\documentclass[final,3p,times,english]{elsarticle}

\usepackage{graphicx}
\usepackage{subfigure}
\usepackage{tikz}
\usepackage{pgfplots}
\usepackage{hyperref}
\hypersetup{
  colorlinks,
  citecolor=Violet,
  linkcolor=Red,
  urlcolor=Blue}
\usepackage{amssymb}
\usepackage{amsmath}

 \usepackage{kpfonts}

\usepackage{multirow}
\usepackage{color}
\usepackage{soul}
\usepackage{multirow}
\usepackage{booktabs}
\usepackage{array}

\usepackage[bold]{hhtensor}

\biboptions{authoryear,semicolon,round}

\usepackage[T1]{fontenc}
\usepackage{charter}
\usepackage[expert]{mathdesign}

\usepackage[latin9]{inputenc}
\usepackage{units}
\usepackage{babel}
\usepackage{sublabel}
\usepackage{tikz}
\usetikzlibrary{arrows,decorations.markings,backgrounds,positioning,fit,shapes,patterns,fadings}
\usepackage{tikz-3dplot}
\usepackage{mathabx}

\usepackage{miller}

\journal{International Journal of Plasticity}

\begin{document}
\newcommand{\etal}{{\it et al.}}

\begin{frontmatter}



\title{Unraveling the temperature dependence of the yield strength in single-crystal tungsten using atomistically-informed crystal plasticity calculations}

\author[add1,add2,add3]{David Cereceda}
\author[add4]{Martin Diehl}
\author[add4]{Franz Roters}
\author[add4]{Dierk Raabe}
\author[add3]{J. Manuel Perlado}
\author[add1]{Jaime Marian\corref{cor1}}
\ead{jmarian@ucla.edu}
\cortext[cor1]{Corresponding author}
\address[add1]{Department of Materials Science and Engineering, University of California Los Angeles, Los Angeles, CA 90095, USA}
\address[add2]{Physical and Life Sciences Directorate, Lawrence Livermore National Laboratory, Livermore, CA, USA}
\address[add3]{Instituto de Fusi\'on Nuclear, Universidad Polit\'ecnica de Madrid, E-28006 Madrid, Spain}
\address[add4]{Max-Planck-Institut f\"ur Eisenforschung, Max-Planck-Stra{\ss}e 1, 40237 D\"usseldorf, Germany}

\begin{abstract}
We use a physically-based crystal plasticity model to predict the yield strength of body-centered cubic (bcc) tungsten single crystals subjected to uniaxial loading. Our model captures the thermally-activated character of screw dislocation motion and full non-Schmid effects, both of which are known to play a critical role in bcc plasticity. The model uses atomistic calculations as the sole source of constitutive information, with no parameter fitting of any kind to experimental data. Our results are in excellent agreement with experimental measurements of the yield stress as a function of temperature for a number of loading orientations. The validated methodology is then employed to calculate the temperature and strain-rate dependence of the yield strength for 231 crystallographic orientations within the standard stereographic triangle. We extract the strain-rate sensitivity of W crystals at different temperatures, and finish with the calculation of yield surfaces under biaxial loading conditions that can be used to define effective yield criteria for engineering design models.

\end{abstract}


\end{frontmatter}
 

\section{Background and motivation}\label{intro}

The plastic behavior of body-centered cubic (bcc) single crystals at low to medium homologous temperatures is governed by the motion of $\frac{1}{2}\langle111\rangle$ screw dislocations on close-packed crystallographic planes. There are two particularities that make bcc metals unique in relation to their deformation characteristics. The first one is the thermally-activated nature of screw dislocation glide, a consequence of the compact (non-planar) structure of the dislocation core at the atomistic level \citep{vitek2004core,gludovatz2011,li2012,samolyuk2013}. This feature is also responsible for the high intrinsic friction stresses reported in the literature for bcc metals and their alloys \citep{romaner2010,samolyuk2013}. The second is the breakdown of the standard geometric projection rule for the resolved shear stress (RSS) from the total stress tensor known as \emph{Schmid law} \citep{schmid1935kristallplastizitat}. This is owed to both specific crystallographic properties of the bcc lattice structure as well as to the coupling between the dislocation core and non-glide components of the stress tensor, which --to the best of our understanding-- is unique to bcc crystals \citep{bulatov1999,brinckmann2008,woodward2001,chaussidon2006,groger2005}. These anomalies have been the subject of much research and discussion going back to the 1960's \citep{takeuchi1967orientation,hull1967orientation,duesbery1969,duesbery_foxall_1969}, both experimentally and --more recently-- using computational atomistic models. 

In regards to the first point above, at low stresses slip proceeds via the thermally activated nucleation of steps on the dislocation line, known as \emph{kink pairs}, and their subsequent sideward relaxation. For a constant strain rate, this gives rise to the characteristic temperature dependence of the flow stress in bcc single crystals, which has been observed for all refractory metals and is considered to be a principal signature of their plastic response \citep{seeger1981,ackermann1983,taylor1992,gordon2010,chaussidon2006,yang2001}. The flow stress is considered to be composed of thermal and athermal contributions, with the latter depending on temperature only as the elastic moduli. Dislocation glide is thought to occur on $\{110\}$, $\{112\}$, and even $\{123\}$ planes, depending on temperature and stress, over a periodic energy landscape known as the \emph{Peierls} potential $U_P$. The connection between the experimentally measured flow stress and this periodic energy potential is via the critical stress for which $U_P$ vanishes at zero temperature, known as the Peierls stress $\sigma_P$. Theoretically then, the flow stress at very low temperatures ($\leq25$ K) is thought to represent the macroscopic equivalent of $\sigma_P$ as the temperature approaches 0 K.
$\sigma_P$ can thus be unequivocally defined and has been the object of considerable numerical work since the first atomistic models were devised by Vitek and co-workers starting in the 1970s \citep{vitek1973core}.

For their part, non-Schmid effects were detected in tests done in the 1930's by Taylor in the wake of his seminal works on plastic flow and strain hardening \citep{taylor1928,taylor1934b,taylor1934a}. Subsequent observations and measurements \citep{sestak_zaburova1965, sherwood1967, zwiesele1979, christian1983, pichl2002,Escaig1968,Escaig1974}, and a rigorous theoretical formulation of the problem \citep{duesbery1998,ito_vitek2001,woodward2001, groger2005, chaussidon2006,groger2008_I,groger2008_II,soare2014plasticity} have established non-Schmid behavior as a principal tenet of bcc plasticity that must be accounted for in order to understand bcc plastic flow. In terms of phenomenology, the two essential aspects to bear in mind are (i) that the resolved shear stress is not independent of the sign of the stress in glide planes of the $\langle111\rangle$ zone (the so-called \emph{twinning/anti-twinning} asymmetry), and (ii) that non-glide components of the stress tensor --{\it i.e.} those which are perpendicular to the Burgers vector-- play a role on the magnitude and sign of the RSS on the glide plane of interest.

Areas where we do not have a complete understanding of bcc plastic picture include the value of the flow stress at near zero absolute temperatures, the meaning of the so-called \emph{knee} temperature, and the onset of athermal flow. In the last two decades, computer simulation has unquestionably emerged as discipline capable of shedding light on these processes on a similar footing with experiments, providing physically-substantiated explanations across a range of temporal and spatial scales. These include the use and application of density-functional theory methods \citep{ventelon2007,ventelon2012,weinberger2013dft,dezerald2014ab,dezerald2015first}, semi empirical atomistic calculations and molecular dynamics calculations \citep{gilbert2011,queyreau2011edge,chang2001,komanduri2001}, kinetic Monte Carlo \citep{lin1999kinetic,cai2002kinetic,deo2002kinetic,scarle2004kinetic,stukowski2015}, and crystal plasticity (CP)\citep{Bassani_NonSchmid_JMPS1991,dao1993,brunig1997}, to name but a few. In general, while there is no doubt that the intricacies associated with $\frac{1}{2}\langle111\rangle$ screw dislocation glide --including its thermally activated nature and deviations from Schmid law-- cannot but be resolved using methods capable of atomistic resolution, one must recognize that, at the same time, flow is a phenomenon potentially involving statistically-significant amounts of dislocations and --as such-- cannot be captured resorting to atomistic calculations only. 

Modeling thermally-activated flow and non-Schmid effects in bcc systems has been the subject of much work, starting in the 1980s and, particularly, in recent times. Different authors have considered different subsets of the $\{110\}$, $\{112\}$, and $\{123\}$ families of glide planes, without \citep{raphanel_bcc_acta1985,holscher1991, holscher1994,raabe1994,Raabe_123slip_1995,raabe1995,Peeters_Acta2000,stainier_bcc_JMPS2002,erieau_bcc_IJP2004, Ma_bcc_CMS2006,hamelin_bcc_IJP2011,kitayama_bcc_IJP2013} and with non-Schmid effects \citep{lee_bcc_IJP1999, kuchnicki_bcc_IJP2008,koester_bcc_acta2012, weinberger_bcc_IJP2012, Alankar_IJP_2013,Lim_MSMSE2013,Narayanan_JMPS_2013, Patra_GeorgiaTech_IJP2014,knezevic_IJP_2014,lim_jmps_2015, lim_ijp_2015}. Of particular interest are some recent simulations where the flow rule is directly formulated on the basis of screw dislocation properties in Fe \citep{yalcinkaya2008,koester_bcc_acta2012,Alankar_IJP_2013,Narayanan_JMPS_2013,Patra_GeorgiaTech_IJP2014,lim_ijp_2015}, Ta \citep{kuchnicki_bcc_IJP2008,Lim_MSMSE2013,knezevic_IJP_2014,lim_jmps_2015}, Mo \citep{yalcinkaya2008,weinberger_bcc_IJP2012,Lim_MSMSE2013,lim_jmps_2015}, W \citep{lee_bcc_IJP1999, weinberger_bcc_IJP2012,Lim_MSMSE2013,knezevic_IJP_2014,lim_jmps_2015}, and Nb \citep{yalcinkaya2008,lim_jmps_2015}. These works also include non-Schmid effects following the model proposed by Vitek and Bassani \citep{duesbery1998,Bassani_NonSchmid_JMPS1991,groger2008_I,groger2008_II}. However, albeit very useful for certain applications, all these works resort to (i) a partial consideration of non-Schmid effects, and (ii) some kind or another of parameter fitting with experimental data, which prevents their use in regions of the parameter space outside the range of fitting and does not link the effective (\emph{macroscopic}) response to exclusively fundamental material properties and features.

In this work, we provide a unified computational methodology consisting of rate-dependent crystal plasticity calculations parameterized entirely and exclusively to atomistic calculations. We show that a full description of non-Schmid effects, together with the state of the art in terms of our understanding of thermally-activated screw dislocation motion, suffices to capture the experimentally measured temperature dependence of the flow stress in tungsten. This is achieved in a fully classical framework, without the need for quantum effects recently invoked to explain the long standing discrepancy observed between the experimentally-measured flow stress below 25 K and calculated values of the Peierls stress \citep{proville2012quantum}. Our methodology also captures the athermal limit of W to within 5\% of the experimental value. We emphasize that this agreement is reached without fitting to any experimental data, all the parameterization is done from first principles atomistic calculations.


Our paper is organized as follows. After this introduction, we provide an overview of the CP method in Section \ref{CP_constitutive_model}. This is followed by Sections \ref{mobb} and \ref{sec:schmid}, where the formulation of the dislocation mobility law and the implementation of non-Schmid effects are presented, including a detailed description of the parameterization procedure employed. The results are given in Section \ref{section_results}, which includes: (i) the validation exercise, with special focus on uniaxial tests as a function of temperature for several loading orientations; (ii) the calculation of temperature and strain rate dependence of the yield strength for uniaxial tensile tests as a function of orientation; and (iii) yield surfaces under biaxial loading conditions as a function of temperature. We finalize in Section \ref{disc} with a brief discussion and the conclusions.

\section{Computational methods}
\subsection{Flow kinematics}\label{flowsec}


\cite{roters2010overview} have presented a detailed review of the kinematic and constitutive aspects of crystal plasticity and here we simply provide a brief overview of the fundamental theory.   
The kinematics for elasto-plastic behavior is defined within the finite deformation framework. The material deformation involves both a reversible lattice response to externally imposed loads or displacements (elastic), and a permanent deformation (irreversible shape change) that remains after all external constraints cease to be applied (plastic). Consequently, crystal plasticity formulations rely on the definition of three reference systems: (i) a fixed coordinate system that represents a laboratory (undeformed) frame of reference, (ii) a current (also known as \emph{material}) frame of reference that represents the global (deformed) shape of the material, and (iii) a \emph{lattice} coordinate system that represents distortions of the underlying crystal structure of the deformed body. Although reference system (i) is used for mathematical convenience, the distinction between (ii) and (iii) is necessary to calculate internal stresses, which arise from distortions defined with respect to a crystallographic frame of reference, as global shape changes may not necessarily have a one-to-one correspondence to internal lattice distortions \citep{lubliner_book,roters2010overview}. 

Mathematically, each point $\vec{X}$ in the reference configuration may be mapped to its image in the current configuration $\vec{x}$ by means of a linear transformation represented by the \emph{deformation gradient} tensor $\matr{F}$, defined as:
\begin{equation}
\matr{F}=\frac{\partial\vec{x}}{\partial\vec{X}}
\label{defgrad}
\end{equation}
In general, $\matr{F}$ is not a symmetric tensor. However, invariance requirements
make it more desirable to work with symmetric measures of strain. One such measure is the so-called \emph{Lagrangian} strain: 
\begin{equation}
\matr{E}=\frac{1}{2}\left(\matr{C}-\matr{I}\right)=\frac{1}{2}\left(\matr{F}^T\matr{F}-\matr{I}\right)
\end{equation}
which refers the deformation of the solid to the reference configuration ($\matr{I}$ is the identity tensor). In the above equation, $\matr{C}$ is the so-called \emph{right Cauchy-Green} tensor.

\begin{figure}[h]
	\centering 
    \begin{tikzpicture}[thick, scale=0.4]

    \pgfmathsetmacro{\xref}{-8}
    \pgfmathsetmacro{\yref}{0}
    
    \pgfmathsetmacro{\xi}{0}
    \pgfmathsetmacro{\yi}{8}
    
    \pgfmathsetmacro{\xf}{8}
    \pgfmathsetmacro{\yf}{0}
    
    \def \delta {2.5}
    \def \epsilon {0.6}
       
        
        \draw  plot[smooth,very thick,tension=.9] coordinates {(-4+\xref,1+\yref) (-3.5+\xref,3+\yref) (-1.5+\xref,4+\yref) (1+\xref,3.5+\yref) (2.5+\xref,3+\yref) (3+\xref,2+\yref)  (3+\xref,-1.5+\yref) (0.5+\xref,-4+\yref) (-3.5+\xref,-2.5+\yref) (-4+\xref,1 +\yref)};
        
         
         \draw  [dashed] plot[smooth,very thick,tension=.9] coordinates {(-4+\xi,1+\yi) (-3.5+\xi,3+\yi) (-1.5+\xi,4+\yi) (1.5+\xi,3.5+\yi) (3+\xi,3+\yi) (3.5+\xi,2+\yi)  (3.5+\xi,-1.5+\yi) (1.0+\xi,-4+\yi) (-3.5+\xi,-2.5+\yi) (-4+\xi,1 +\yi)};
        
        
        \draw  plot[smooth,very thick,tension=.9] coordinates {(-4+\xf+1.2*\epsilon,1+\yf) (-3.5+\xf+1.8*\epsilon,3+\yf) (-1.5+\xf+2.1*\epsilon,4+\yf) (1.5+\xf+2.0*\epsilon,3.5+\yf) (3+\xf+1.8*\epsilon,3+\yf) (3.5+\xf+1.5*\epsilon,2+\yf)  (3.5+\xf+0.66*\epsilon,-1.5+\yf) (1.0+\xf+0*\epsilon,-4+\yf) (-3.5+\xf+0.33*\epsilon,-2.5+\yf) (-4+\xf+1.2*\epsilon,1 +\yf)};
        
        \node (A) at (0,-1) {\Large $\matr{F}$};
        \node (B) at (\xref*1.05,\yi*0.7) {\Large $\matr{F}_p$};
        \node (C) at (\xf*1.05,\yi*0.7) {\Large $\matr{F}_e$};
        \draw[black,very thick,->] (\xref+\delta,\yref) -- (\xf-\delta,\yf);
        \draw (\xref,\yref+\delta) node[circle, inner sep=0.8pt, fill=white] (C) {};  
        \draw (\xi-\delta,\yi) node[circle, inner sep=0.8pt, fill=white] (D) {};  
        \draw [very thick,->] (C) to [bend left=45] (D); 
        \draw (\xi+\delta,\yi) node[circle, inner sep=0.8pt, fill=white] (E) {};  
        \draw (\xf,\yf+\delta) node[circle, inner sep=0.8pt, fill=white] (F) {};  
        \draw [very thick,->] (E) to [bend left=45] (F); 
        
        
        \node (D) at (\xi*1.05,\yi-2.6) {\it Intermediate};
        \node (E) at (\xref*1.05,\yi-10.8) {{\it Reference}};
        \node (F) at (\xf*1.05,\yi-10.8) {{\it Material}};

\node[circle,shading=ball,minimum width=0.3cm] (ball) at (\xref-2,\yref-1.5) {};
\node[circle,shading=ball,minimum width=0.3cm] (ball) at (\xref-2,\yref-0.5) {};
\node[circle,shading=ball,minimum width=0.3cm] (ball) at (\xref-2,\yref+0.5) {};
\node[circle,shading=ball,minimum width=0.3cm] (ball) at (\xref-2,\yref+1.5) {};
	
\node[circle,shading=ball,minimum width=0.3cm] (ball) at (\xref-1,\yref-1.5) {};
\node[circle,shading=ball,minimum width=0.3cm] (ball) at (\xref-1,\yref-0.5) {};
\node[circle,shading=ball,minimum width=0.3cm] (ball) at (\xref-1,\yref+0.5) {};
\node[circle,shading=ball,minimum width=0.3cm] (ball) at (\xref-1,\yref+1.5) {};

\node[circle,shading=ball,minimum width=0.3cm] (ball) at (\xref,\yref-1.5) {};
\node[circle,shading=ball,minimum width=0.3cm] (ball) at (\xref,\yref-0.5) {};
\node[circle,shading=ball,minimum width=0.3cm] (ball) at (\xref,\yref+0.5) {};
\node[circle,shading=ball,minimum width=0.3cm] (ball) at (\xref,\yref+1.5) {};
	
\node[circle,shading=ball,minimum width=0.3cm] (ball) at (\xref+1,\yref-1.5) {};
\node[circle,shading=ball,minimum width=0.3cm] (ball) at (\xref+1,\yref-0.5) {};
\node[circle,shading=ball,minimum width=0.3cm] (ball) at (\xref+1,\yref+0.5) {};
\node[circle,shading=ball,minimum width=0.3cm] (ball) at (\xref+1,\yref+1.5) {};
%
		
   
\node[circle,shading=ball,minimum width=0.3cm] (ball) at (\xi-2,\yi-1.5) {};
\node[circle,shading=ball,minimum width=0.3cm] (ball) at (\xi-2,\yi-0.5) {};
\node[circle,shading=ball,minimum width=0.3cm] (ball) at (\xi-1,\yi+0.5) {};
\node[circle,shading=ball,minimum width=0.3cm] (ball) at (\xi-1,\yi+1.5) {};
 
\node[circle,shading=ball,minimum width=0.3cm] (ball) at (\xi-1,\yi-1.5) {};
\node[circle,shading=ball,minimum width=0.3cm] (ball) at (\xi-1,\yi-0.5) {};
\node[circle,shading=ball,minimum width=0.3cm] (ball) at (\xi,\yi+0.5) {};
\node[circle,shading=ball,minimum width=0.3cm] (ball) at (\xi,\yi+1.5) {};

\node[circle,shading=ball,minimum width=0.3cm] (ball) at (\xi,\yi-1.5) {};
\node[circle,shading=ball,minimum width=0.3cm] (ball) at (\xi,\yi-0.5) {};
\node[circle,shading=ball,minimum width=0.3cm] (ball) at (\xi+1,\yi+0.5) {};
\node[circle,shading=ball,minimum width=0.3cm] (ball) at (\xi+1,\yi+1.5) {};

\node[circle,shading=ball,minimum width=0.3cm] (ball) at (\xi+1,\yi-1.5) {};
\node[circle,shading=ball,minimum width=0.3cm] (ball) at (\xi+1,\yi-0.5) {};
\node[circle,shading=ball,minimum width=0.3cm] (ball) at (\xi+2,\yi+0.5) {};
\node[circle,shading=ball,minimum width=0.3cm] (ball) at (\xi+2,\yi+1.5) {};

%
		
   
\node[circle,shading=ball,minimum width=0.3cm] (ball) at (\xf-2,\yf-1.5) {};
\node[circle,shading=ball,minimum width=0.3cm] (ball) at (\xf-2+0.33*\epsilon,\yf-0.5) {};
\node[circle,shading=ball,minimum width=0.3cm] (ball) at (\xf-1+0.66*\epsilon,\yf+0.5) {};
\node[circle,shading=ball,minimum width=0.3cm] (ball) at (\xf-1+\epsilon,\yf+1.5) {};
 
\node[circle,shading=ball,minimum width=0.3cm] (ball) at (\xf-1,\yf-1.5) {};
\node[circle,shading=ball,minimum width=0.3cm] (ball) at (\xf-1+0.33*\epsilon,\yf-0.5) {};
\node[circle,shading=ball,minimum width=0.3cm] (ball) at (\xf+0.66*\epsilon,\yf+0.5) {};
\node[circle,shading=ball,minimum width=0.3cm] (ball) at (\xf+\epsilon,\yf+1.5) {};

\node[circle,shading=ball,minimum width=0.3cm] (ball) at (\xf,\yf-1.5) {};
\node[circle,shading=ball,minimum width=0.3cm] (ball) at (\xf+0.33*\epsilon,\yf-0.5) {};
\node[circle,shading=ball,minimum width=0.3cm] (ball) at (\xf+1+0.66*\epsilon,\yf+0.5) {};
\node[circle,shading=ball,minimum width=0.3cm] (ball) at (\xf+1+\epsilon,\yf+1.5) {};

\node[circle,shading=ball,minimum width=0.3cm] (ball) at (\xf+1,\yf-1.5) {};
\node[circle,shading=ball,minimum width=0.3cm] (ball) at (\xf+1+0.33*\epsilon,\yf-0.5) {};
\node[circle,shading=ball,minimum width=0.3cm] (ball) at (\xf+2+0.66*\epsilon,\yf+0.5) {};
\node[circle,shading=ball,minimum width=0.3cm] (ball) at (\xf+2+\epsilon,\yf+1.5) {};

        
     \end{tikzpicture}
        
    \caption{Multiplicative decomposition of the deformation gradient $\matr{F}$. }
	\label{kin}
\end{figure}
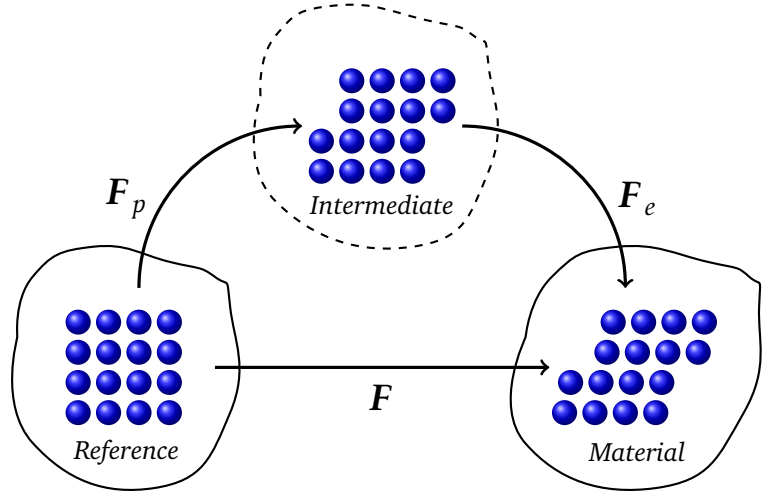

Following \citet{lee1969}, the total deformation gradient $\matr{F}$ can be multiplicatively decomposed into an elastic, $\matr{F}_e$, and a plastic, $\matr{F}_p$, part\footnote{It must be noted that other decompositions are also admissible \citep{fish2000finite}. The reader is referred to the work by \citet{reina_JMPS2014} for a discussion on the uniqueness and validity of the multiplicative decomposition.}, {\it i.e.}:
\begin{equation}
\matr{F}=\matr{F}_e\matr{F}_p
\label{mult}
\end{equation}
whence
$$\matr{F}_e=\matr{F}\matr{F}_p^{-1}\Leftrightarrow \matr{F}_p=\matr{F}_e^{-1}\matr{F}$$
This is schematically shown in Figure \ref{kin}, where the relationship between the reference, intermediate, and current configurations is provided.
To close the CP model, we take the time rate in eq.\ \eqref{defgrad}, which results in:
\begin{equation}
\matr{L}=\matr{\dot{F}}\matr{F}^{-1}=\matr{\dot{F}}_e\matr{F}_e^{-1}+\matr{F}_e\left(\matr{\dot{F}}_p\matr{F}^{-1}_p\right)\matr{F}_e^{-1}=\matr{L}_e+\matr{F}_e\matr{L}_p\matr{F}_e^{-1}
\end{equation}
where $\matr{L}_p$ is the plastic velocity gradient, which is evaluated in the intermediate configuration and must therefore be mapped into the current configuration by $\matr{F}_e$. Constitutive information enters the CP model via $\matr{L}_p$, which is described in the following section. 

The above finite-deformation kinematic frameworkis implemented into the D\"usseldorf Advanced Materials Simulation Kit (DAMASK), which is the tool employed in this work to carry out of the calculations. DAMASK is a flexible and hierarchically structured model of material point behavior for the solution of elastoplastic boundary value problems along with damage and thermal physics \citep{DAMASK}.

\subsection{Solution procedure and constitutive model}
\label{CP_constitutive_model}

A \emph{Hookean} constitutive response is assumed such that the stress depends linearly on the elastic strain via the anisotropic elastic stiffness tensor $\matr{{\mathbb C}}$. Both the stress and strain measures that are used internally are formulated in terms of material coordinates. 
For the stress, we use the \emph{second Piola-Kirchhoff} stress measure $\matr{S}$, defined as:
\begin{equation}
\matr{S}=\matr{\mathbb{C}}:\matr{E}_e=\frac{\matr{\mathbb{C}}}{2}\left(\matr{F}_e^T\matr{F}_e-\matr{I}\right)
\label{pk2}
\end{equation}
where $\matr{E_e}$ is the (elastic) \emph{Green-Lagrange} strain tensor.
$\matr{S}$ and $\matr{E}_e$ are both symmetric material tensors, and thus $\matr{{\mathbb C}}$ is itself symmetric such that a general $3\times3\times3\times3$ tensor can be written as a $6\times6$ matrix. For cubic lattices, $\matr{{\mathbb C}}$ can be reduced by symmetry to only the three independent elastic constants ${\mathbb C}_{11}$, ${\mathbb C}_{12}$, and ${\mathbb C}_{44}$.

The stress $\matr{S}$ acts as the driving force for the plastic velocity gradient $\matr{L}_p$. $\matr{L}_p$ depends on the underlying microstructure via a set of state variables $\matr{\xi}$ defined by the plasticity model employed:
\begin{equation}
\matr{L}_p=f(\matr{S},\matr{\xi},\ldots)
\label{lp}
\end{equation}
$\matr{L}_p$ controls the evolution of the plastic deformation gradient:
\begin{equation}
\matr{\dot{F}}_p=\matr{L}_p\matr{F}_p
\label{fplp}
\end{equation}
The set of nonlinear eqs.\ \eqref{mult} and \eqref{pk2} to \eqref{fplp} must be solved iteratively, which in DAMASK is done by using an integration algorithm based on the implicit scheme originally proposed by \citet{Kalididndi1992}. 
The linear system is solved iteratively using the Newton-Raphson technique and, once convergence is achieved, the plastic deformation gradient is obtained using the Euler backward update.
In this integration scheme described above, the primary variable to solve for is the plastic velocity gradient. However, one may devise schemes where the primary variables are the stress, the plastic or elastic deformation gradient, the internal variables or a combination thereof. Such schemes may be chosen on the basis of computational efficiency \citep{dumoulin2009}.

Constitutive information for the plastic regime enters the CP model via eq.\ \eqref{lp}, where the dependencies of the flow rule on each of the state variables are established.  It is here where the plastic deformation modes are defined, their geometric particularities, as well as specifics associated with the crystal structure under study.  The CP model must also include evolution equations for the state variables $\xi$:
\begin{equation}
\matr{\dot\xi}=g(\matr{S},\matr{\xi},\ldots)
\label{xi}
\end{equation}
where the details again depend on the model selected. In DAMASK, various integration schemes for the state update exist \citep{DAMASK}. Then, two integration schemes are performed staggered: eqs.\ \eqref{mult} to \eqref{fplp} are solved at a fixed plastic state, followed by a state update. This procedure is iteratively repeated until a converged solution is achieved within the given tolerances. More details about the implementation of this technique in the code are given by \citet{Kalididndi1992}.
In general then, the stress in the CP model can be considered a response function of the position $\vec{r}$, the deformation state $\matr{F}$, the set of state variables $\xi$, and a set of boundary conditions, {\it i.e.}\:
\begin{equation}
\matr{S} = f(\vec{r},\matr{F},\vec{\xi},\ldots)
\label{fftt}
\end{equation}

In these calculations we are interested in simulating engineering stress-strain tests, and --consequently-- it is helpful to express the results in the reference coordinate frame. For the stress, we use the \emph{first} Piola-Kirchhoff measure defined as:
$$\matr{P}=\matr{F}_e\matr{S}$$
Note that, although in general $\matr{P}$ is not symmetric, for uniaxial tension tests $\matr{P}$ is a symmetric tensor on account of $\matr{F}_e$ being symmetric. For the strain, we use the \emph{Biot} tensor:
$$\matr{B}=\matr{U}-\matr{I}=\sqrt{\matr{F}^T\matr{F}}-\matr{I}$$
where $\matr{U}$ is the \emph{right stretch} tensor\footnote{$\matr{U}$ emerges from the so-called \emph{polar} decomposition: $\matr{F}=\matr{R}\matr{U}$, where $\matr{R}$ is a matrix the represents rigid rotation, and $\matr{U}$ is a pure stretch. Plasticity-induced crystal rotations are very important and give rise to crystallographic texture evolution in deformed crystals. However, only yielding is of concern here, and thus $\matr{U}$ is the component of interest.}. The stress-strain curves shown throughout this paper are obtained by tracking the evolution of $P_{zz}$ and $B_{zz}$, where $z$ is designated as the loading direction.
For uniaxial loading simulations, $\matr{\dot{F}}\equiv\matr{\dot{F}}^T$ and thus $\matr{\dot{C}}=\matr{\dot{F}}\approx\matr{\dot{B}}$. We refer to the deformation rate represented by $\dot{F}_{zz}$ generically as $\dot{\varepsilon}$ in the remainder of this paper.

\subsection{The flow rule}

In the present CP calculations it is assumed that all the plastic deformation is due to dislocation slip. Then, the plastic velocity gradient can be written as:
\begin{equation}
\matr{L}_p=\sum_{\alpha}\matr{P}_{\rm S}^{\alpha}\dot{\gamma}^{\alpha}
\label{lpp}
\end{equation}
where $\dot{\gamma}^{\alpha}$ is the slip rate on slip system $\alpha$, and $\matr{P}_{\rm S}^{\alpha}$ is a geometric projection tensor that will be defined later.
The slip rate is calculated from Orowan's equation:
\begin{equation}
\dot{\gamma}^{\alpha}=b\rho^{\alpha} v_s(\tau^{\alpha},T)
\label{orowan}
\end{equation}
where $b=a_0\sqrt{3}/2$ is the modulus of the Burgers vector, $a_0$ is the lattice parameter, $T$ the absolute temperature, $\rho^{\alpha}$ is the (mobile) screw dislocation density in slip system ${\alpha}$, and $v_s(\tau^{\alpha},T)$ is the screw dislocation velocity. The present formulation of the flow rule belongs to the class of \emph{non-associated}, \emph{rate-dependent} CP models \citep{mcdowell2008}.

The two characteristics that are particular to bcc plasticity are the thermally-activated nature of screw dislocation motion, which makes it the rate-controlling process during plastic deformation, and the existence of non-Schmid effects, {\it i.e.} deviations from the geometric projection law for the resolved shear stress. Both of these physical processes have been known for several decades, and have been carefully analyzed experimentally (cf.\ Section \ref{intro}). If our intent is to predict the temperature dependence of the flow stress in bcc metals, accurate physical descriptions of both of them must be incorporated into our CP model. This is the subject of the following sections. As we shall see, non-Schmid effects establish the form of the projection tensor $\matr{P}_{\rm tot}^{\alpha}$, while the velocity $v_s(\tau^{\alpha},T)$ captures the thermally activated character of dislocation motion. We make tungsten the object of our study for a number reasons presented in previous works \citep{Cereceda_Wpotentials,stukowski2015}\footnote{W is an elastic isotropic metal, which simplifies the constitutive plastic formulation.}.

\subsubsection{Screw dislocation mobility law}
\label{mobb}
Except at high homologous temperatures and strain rates, screw dislocation motion is the rate-limiting step in bcc crystal deformation. Although recent dislocation dynamics simulations in $\alpha$-Fe challenge the notion that the dislocation density is monolithic across the entire temperature range \citep{Monnet2009174,Naamane201084,Monnet2011451,tang2014}, it is reasonable to assume a dominance of screw dislocations in the temperature and strain rate regimes considered in this work ($0<T/T_m<0.2$ and $\dot{\varepsilon}\approx10^{-4}$ s$^{-1}$). In the thermally activated regime, screw dislocation motion proceeds via the nucleation of kink-pairs and their subsequent lateral relaxation. In the regime of interest here, kink relaxation is a significantly faster process than kink-pair nucleation, and it can thus be assumed that no new kink-pairs will be nucleated while lateral kink motion is underway \citep{stukowski2015}. 
Such assumption leads to the following expression for the total time, $t_t$, required for a kink pair to form and sweep a rectilinear screw dislocation segment of length $\lambda^{\alpha}$ lying on a given slip plane:
\begin{equation}
t_t=t_n+t_k=J(\tau^{\alpha},T)^{-1}+\frac{\lambda^{\alpha}-w}{2v_k(\tau^{\alpha},T)}
\label{monkey1}
\end{equation}
where $t_n$ is the mean time to nucleate a kink pair, $t_k$ is the time needed for a kink to sweep half a segment length, $J$ is the kink-pair nucleation rate, $w$ is the kink-pair separation, and $v_k$ is the kink velocity. 
The kink-pair nucleation rate follows an Arrhenius formulation \citep{stukowski2015}:
\begin{equation}
J(\tau^{\alpha},T)=\frac{\nu_0(\lambda^{\alpha}-w)}{b}\exp{\left(-\frac{\Delta H_{kp}(\tau^{\alpha})}{kT}\right)}
\label{j}
\end{equation}
where $\nu_0$ is an attempt frequency, $\Delta H_{kp}$ is the activation enthalpy of a kink pair at stress $\tau^{\alpha}$, and $k$ is Boltzmann's constant. For its part, the kink velocity can be expressed as \citep{dorn1964,kocks_argon_ashby_1975}:
\begin{equation}
v_k(\tau^{\alpha},T)=\frac{b\tau^{\alpha}}{B(T)}
\label{vk}
\end{equation}
where $B$ is friction coefficient typically assumed to be linearly dependent on temperature. However, calculations made to obtain the value of $B$ for the interatomic potential employed in this work, have yielded no temperature dependence, and here $B$ is simply a constant \citep{swinburne2013theory} .
The dislocation velocity can be obtained after operating with eqs.\ \eqref{j} and \eqref{vk} as: 
\begin{equation}
v_s=\frac{h}{t_t}=\frac{h}{t_n+t_k}=
\frac{2bh\tau^{\alpha}\nu_0(\lambda^{\alpha}-w)\exp{\left(-\frac{\Delta H_{kp}}{kT}\right)}}{
2b^2\tau^{\alpha}+\nu_0B(\lambda^{\alpha}-w)^2\exp{\left(-\frac{\Delta H_{kp}}{kT}\right)}
}\label{mob1}
\end{equation}
where $h=a_0\sqrt{6}/3$ is the distance between two consecutive Peierls valleys. We note that at low temperatures, or when $t_k\ll t_n$, the second term in the denominator vanishes and one recovers the standard diffusive velocity expression  commonly used in crystal plasticity and dislocation dynamics:
$$v_s=\nu_0 h\frac{(\lambda^{\alpha}-w)}{b}\exp{\left(-\frac{\Delta H_{kp}(\tau^{\alpha})}{kT}\right)}$$

\begin{figure}[h]
\includegraphics[width=1.0\linewidth]{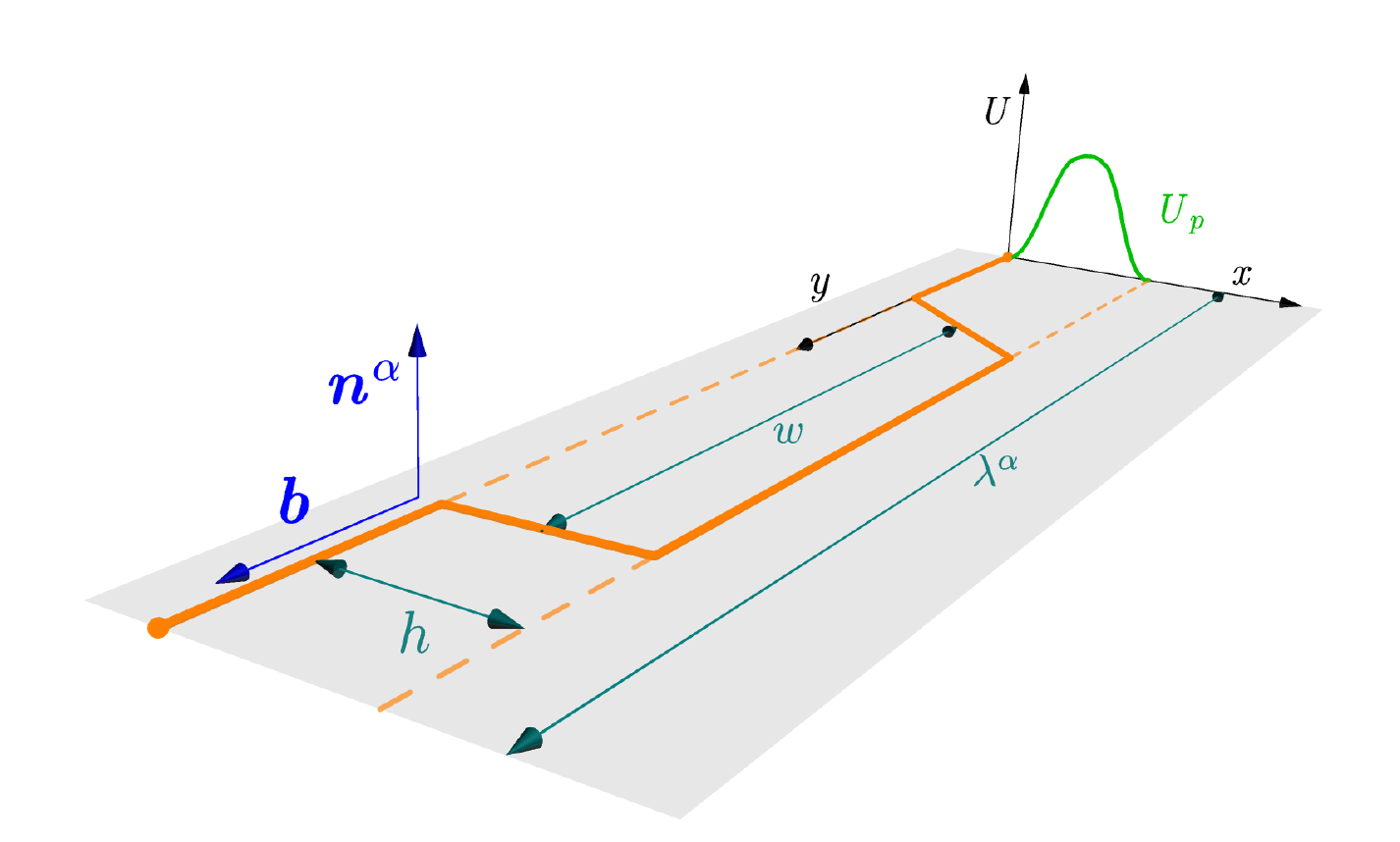}
\caption{\label{kinks-tikz}Schematic depiction of a kink pair on a screw segment of length $\lambda$ lying on a slip plane $\vec{n}^\alpha$ (of the $\{110\}$ family). The vertical axis represents the potential energy, with the Peierls potential clearly marked. The dashed line represents the initial equilibrium line position.} 
\end{figure} 

The parameterization of eq.\ \eqref{mob1} is a critical step that establishes a physical connection with the scales where kink-pairs are resolved as atomistic entities. This is the first essential piece of physics required to achieve predictive capabilities. We have devoted much effort in past works to calculate the necessary parameters from fundamental models based on semiempirical interatomic potentials \citep{Cereceda_Wpotentials,stukowski2015}. 
The list of parameters employed in this work and their associated values and units are given in Table \ref{table2}. The physical meaning of some of these parameters is best expressed in pictorial form.
Figure \ref{kinks-tikz} shows a schematic diagram of the topology of a kink pair lying on the Peierls energy substrate. The figure highlights the physical meaning of each parameter listed in the table. In addition to the references provided earlier, a detailed description of the protocols used to calculate all the adjustable parameters in our formulation is provided by \citet{cereceda2015multiscale}.

At this stage, it is worth to introduce a note about the available slip systems (which establish the running indices of $\alpha$. \citet{stukowski2015} have shown that in W an elementary glide on a $\{112\}$ plane is a composite of two elementary steps on alternate $\{110\}$ planes. Judging by these results, we conclude that glide on any given plane is achieved by way of sequential $\{110\}$ jumps, which constitutes the basis to simulate plastic yielding in the foregoing Sections. This is consistent with recent atomistic simulations \citep{Cereceda_Wpotentials} and experiments \citep{caillard2010_I,caillard2010_II,marichal2013,marichal2014} and limits the number of available slip systems in our study to 12 (listed in \ref{app:slip}). We note that this model of slip for W is not necessarily suggestive of what may happen in other bcc crystals \citep{Franciosi2015226}.

\subsubsection{Projection tensor and non-Schmid effects}
\label{sec:schmid}

The tensor $\matr{P}_{\rm S}^{\alpha}$ introduced in eq.\ \eqref{lpp} represents the Schmid (geometric) projection of the strain rate contribution from a slip system defined by the plane normal $\matr{n}^{\alpha}$ and slip direction $\matr{m}^{\alpha}$ (both unit vectors). However, as pointed out above, $\matr{P}_{\rm S}^{\alpha}$ does not capture the full panoply of non-Schmid effects needed to calculate the value of the resolved shear stress on that slip system, $\tau^{\alpha}$. For this, we introduce a total projection tensor $\matr{P}_{\rm tot}^{\alpha}$ such that:
\begin{equation}
\tau^{\alpha}=\matr{P}_{\rm tot}^{\alpha}:\matr{\sigma}=\left(\matr{P}_{\rm S}^{\alpha}+\matr{P}_{\rm T/AT}^{\alpha}+\matr{P}_{\rm ng}^{\alpha}\right):\matr{\sigma}
\label{pepe}
\end{equation}
where 
\begin{equation}
\matr{P}_{\rm S}^{\alpha}=\vec{m}^{\alpha}\otimes\vec{n}^{\alpha}
\end{equation}
is the Schmid tensor, with 
$$\matr{\sigma}=J^{-1}\matr{F}\matr{S}\matr{F}^T$$
the \emph{Cauchy} (true) stress and $J=\det({\matr{F}})$ the \emph{Jacobian}. The tensors
\begin{eqnarray}
&\matr{P}_{\rm T/AT}^{\alpha}&=a_1\vec{m}^{\alpha}\otimes\vec{n}_1^{\alpha}\\
&\matr{P}_{\rm ng}^{\alpha}&=a_2\left(\vec{n}^{\alpha}\times\vec{m}^{\alpha}\right)\otimes\vec{n}^{\alpha}+a_3\left(\vec{n}_1^{\alpha}\times\vec{m}^{\alpha}\right)\otimes\vec{n}_1^{\alpha}
\end{eqnarray}
are non-Schmid tensors representing respectively the twinning/anti-twinning asymmetry (T/AT) and the effects due to non-glide stress components. $a_1$, $a_2$, and $a_3$ are material-dependent constants that must also be calculated and added to our parameterization database. The vector $\vec{n}_1^{\alpha}$ forms an angle of $-60^{\circ}$ with the reference slip plane defined by $\vec{n}^{\alpha}$, and changes sign with the direction of slip on each glide plane \citep{koester_bcc_acta2012}.

\begin{figure}[h]
\centering
\includegraphics[width=1.0\linewidth]{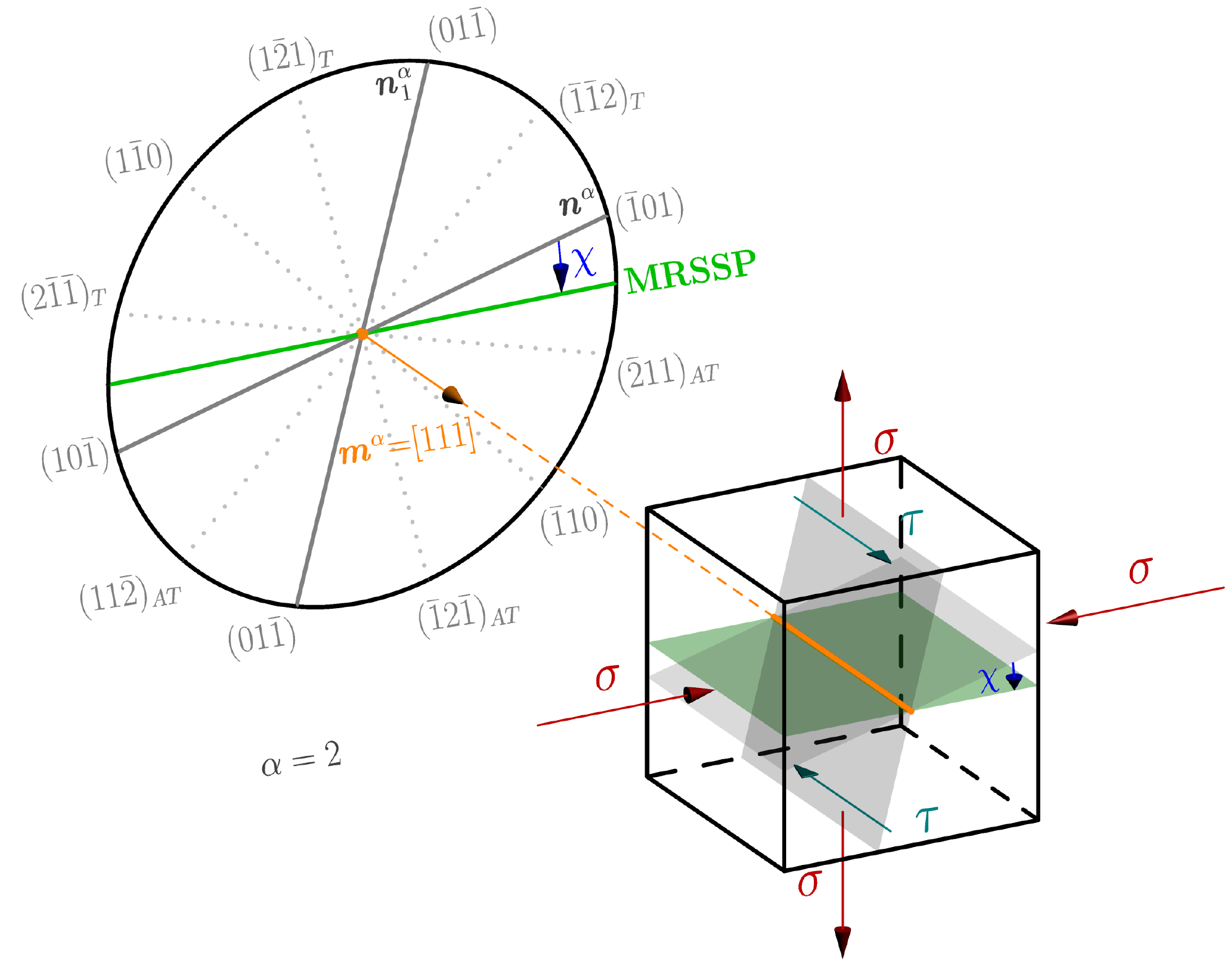}
\caption{\label{davinci}Crystallographic diagram of the [111] zone in the bcc lattice with each \{110\} and \{112\} clearly labeled. The picture also shows a mapping of the [111] zone to a schematic atomistic box containing a screw dislocation subjected to shear and nonglide stresses according to Vitek's convention. This setup is used to calculate the critical RSS using atomistic calculations (cf.\ \ref{app_nonschmid}). The glide $\vec{n}^{\alpha}$, auxiliary $\vec{n_1}^{\alpha}$ and MRSS planes are labeled in each case. A $[\bar{1}01]$ glide plane corresponds to $\alpha=2$ in our CP calculations.}
\end{figure}

\begin{figure}[h]
\centering
\includegraphics[width=1.0\linewidth]{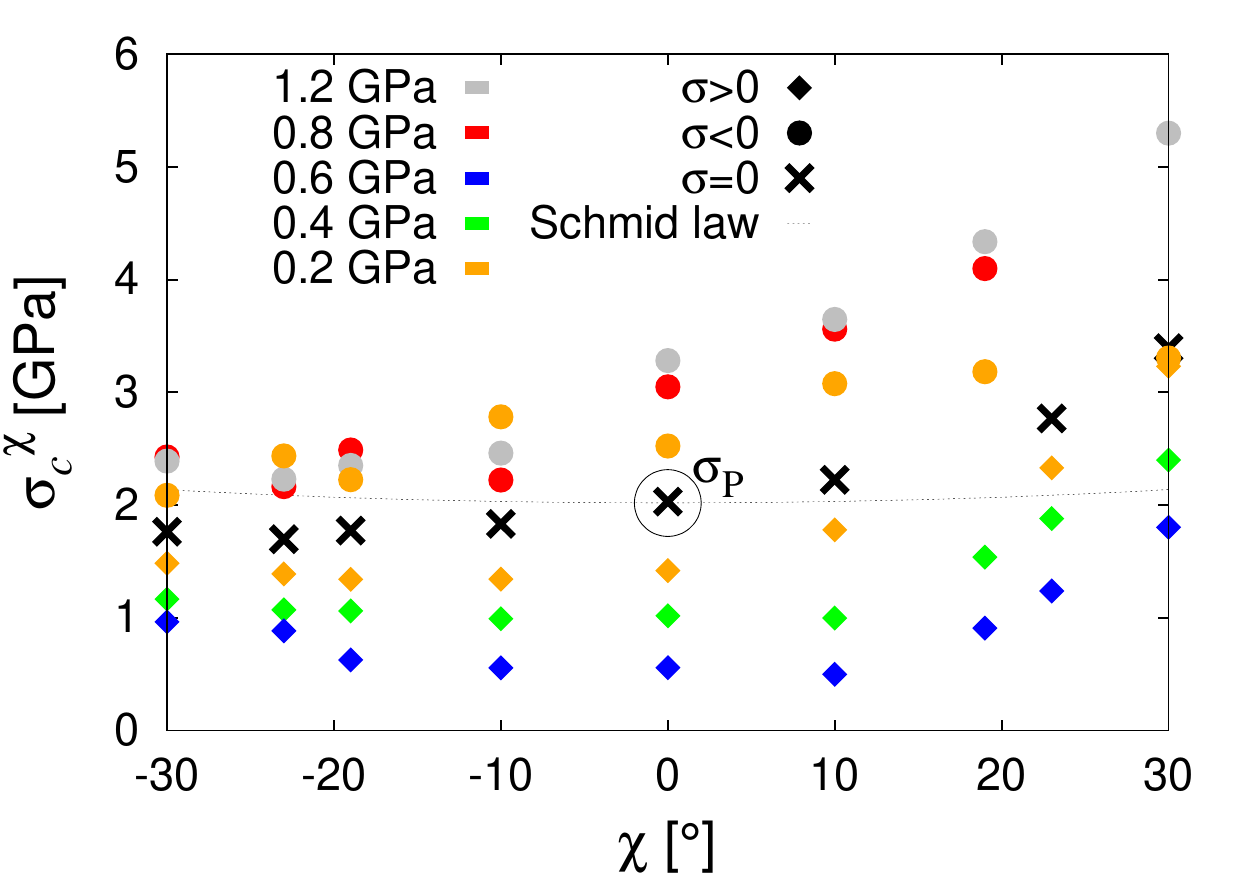}
\caption{\label{non}Critical resolved shear stress as a function of the angle $\chi$ between the MRSS and glide planes and the value of the nonglide stress component $\sigma$ with the sign convention according to Fig.\ \ref{davinci}. The value of the Peierls stress $\sigma_P=2.03$ GPa is circled.} 
\end{figure}

The present non-Schmid formulation was originally developed by Vitek and expanded by others, and has been successfully used to propose yielding criteria adapted to finite element and crystal plasticity calculations in a number of cases \citep{groger2008_III,chen2013_Gumbsch,weinberger_bcc_IJP2012}. The reader is referred to these works for more details but it is worth pointing out that the methodology that these authors have proposed is not unique, and that other rigorous implementations of non-Schmid effects could equally be devised. 
For the purposes of this section, suffice it to say that the particularities of the screw dislocation core and the bcc lattice structure result in deviations from a purely geometric projection. These deviations originate, respectively, from a geometric asymmetry between the twinning and anti-twinning directions of the $\langle111\rangle$ zone --from which $a_1$ is first calculated--, and from the effect that nonglide components (termed generically `$\sigma$') of the local stress tensor have on the critical resolved shear stress, from which $a_2$ and $a_3$ are obtained.  Atomistic calculations specifically designed to calculate the non-Schmid critical stress $\tau_c^{\chi}$ as a function of the angle $\chi$ between the maximum resolved shear stress (MRSS) plane were performed according to the geometry shown schematically in Figure \ref{davinci}. The Figure shows the mapping between the atomistic box and the crystallography of the $[111]$ zone. Following the sign convention used in the Figure, the stress tensor applied is:
\begin{equation}
\left( \begin{array}{ccc}
-\sigma & 0 & 0 \\
0 & \sigma & \tau \\
0 & \tau & 0 \end{array} \right)
\label{tensor}
\end{equation}
which activates axial (nonglide) stress components while maintainign zero pressure.
$\tau_c^{\chi}$ is expressed as a combination of the contributions displayed in Fig.\ \ref{davinci}:
\begin{equation}
\tau_c^{\chi}=\frac{\tau_c^{\ast}+\sigma\left(a_2\sin(2\chi)+a_3\sin\left(2\chi+\frac{\pi}{6}\right)\right)}{\cos\chi + a_1\cos\left(\chi+\frac{\pi}{3}\right)}
\label{tauc}
\end{equation}
where $\tau_c^{\ast}$ is a fitting constant that represents the Peierls stress.
The details of these atomistic calculations are provided in \ref{app_nonschmid}.  The results for $\tau_c^{\chi}$ are shown in Figure \ref{non} as a function of $\chi$ and $\sigma$, with $\tau^{\ast}_c$, $a_1$, $a_2$, and $a_3$ given in Table \ref{table2}. It is worth noting that the relation between $\tau_c^{\chi}$ and $\sigma$ has been established for tensile nonglide stresses only ($\sigma>0$), for consistency with the linear dependence used in the work of Vitek and collaborators \citep{groger2008_I,groger2008_II} that has been used in other crystal plasticity works \citep{koester_bcc_acta2012}. However, nothing precludes the use of nonlinear fitting functions that capture both the tensile and compressive regimes simultaneously (cf.\ \ref{app_nonschmid}). It is worth noting that \citet{groger2008_II} obtained values of $a_1=0$, $a_2=0.56$, and $a_3=0.75$ using a bond-order potential, substantially far from our values for those parameters.

\begin{figure}[h]
\centering
\includegraphics[width=.75\linewidth]{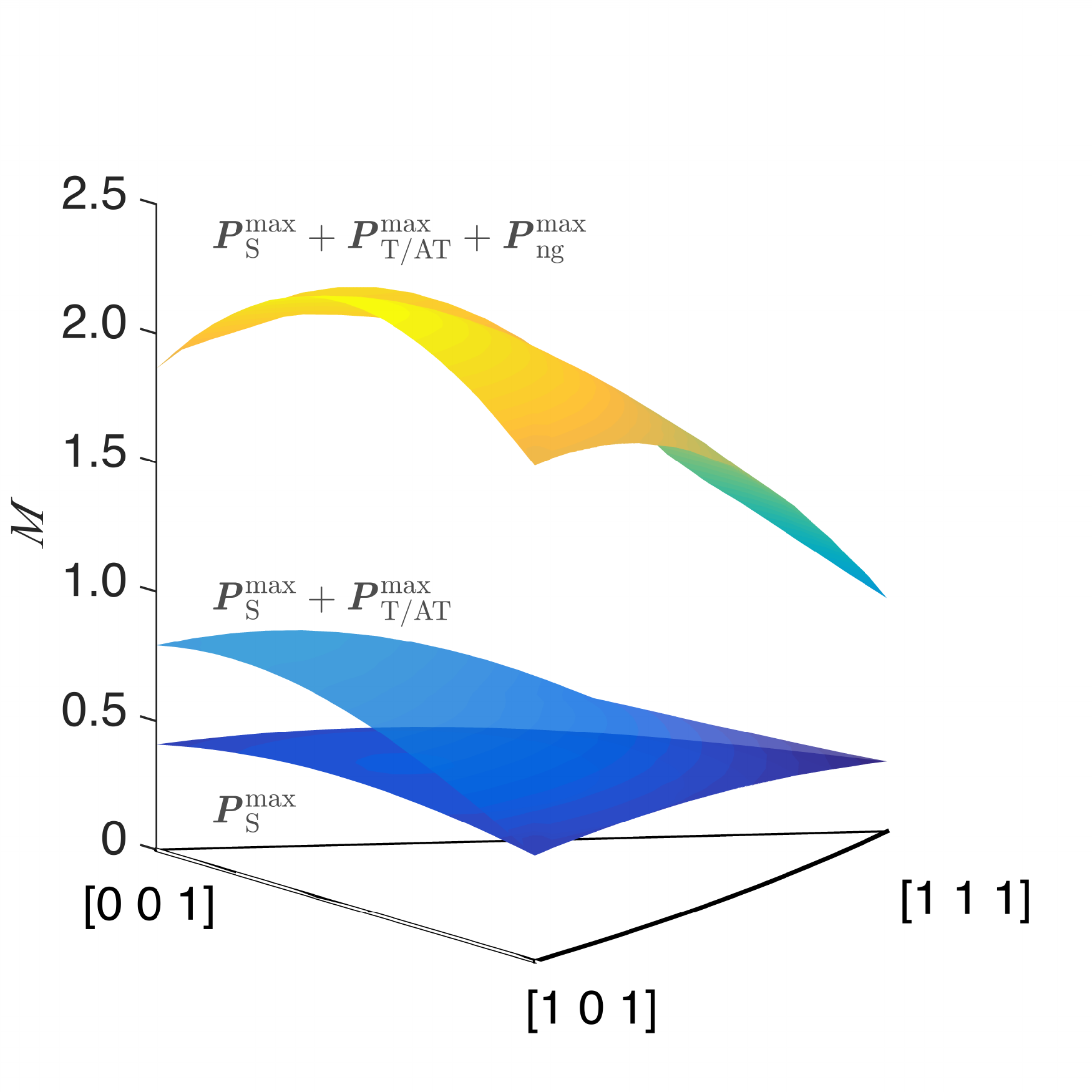}
\caption{\label{factor}Projection factor according to eq.\ \eqref{mmm} for 231 directions within the standard triangle. The contributions of each of therms in eq.\ \eqref{pepe} are broken down for comparison.} 
\end{figure}

By way of example, we calculate the maximum projection factor $M$ for directions in the standard stereographic triangle using the fully parameterized projection tensor:
\begin{equation}
M=(\vec{l}\otimes\vec{l}):\matr{P}_{\rm tot}=(\vec{l}\otimes\vec{l}):(\matr{P}_{\rm S}^{\alpha}+\matr{P}_{\rm T/AT}^{\alpha}+\matr{P}_{\rm ng}^{\alpha})
\label{mmm}
\end{equation}
where $\vec{l}$ is the loading direction, which is obtained by visiting each of the nodes resulting from the discretization of the standard triangle area into a uniform grid consisting of 231 points. The results for tension ($\sigma>0$) are shown in Figure \ref{factor}. It is clear than non-Schmid effects --particularly the impact of nonglide components-- are critical to calculate the RSS on a given slip system. We find that from a maximum nominal value of $M=0.5$ for the standard Schmid law ($\matr{P}_{\rm S}^{\rm max}$) there is a twofold amplification when the twinning/anti-twinning asymmetry is considered ($\matr{P}_{\rm S}^{\rm max}+\matr{P}_{\rm T/AT}^{\rm max}$), and an astonishing fourfold increase when nonglide effects are also included ($\matr{P}_{\rm S}^{\rm max}+\matr{P}_{\rm T/AT}^{\rm max}+\matr{P}_{\rm ng}^{\rm max}$). As we shall see in Section \ref{validation}, this has extraordinary importance when comparing CP calculations to experimental measurements.

\subsection{Dislocation density evolution model}
\label{evol}
To close the model, one needs to provide an evolution law for the dislocation density in Orowan's equation \ref{orowan}.
There are numerous density evolution models proposed in the literature, each with a specific domain of applicability \citep{mecking1981,estrin1996dislocation,Arsenlis20021979,stainier_bcc_JMPS2002,barton2011}. In this work we are mainly interested in yielding, {\it i.e.} the elastic-to-plastic transition before dislocation-based slip takes on a dominant role in the constitutive model. We use the model presented by \citet{Franz_Roters_thesis}, in which the mobile dislocation density on slip system $\alpha$ evolves in time according to:
\begin{equation}
\dot{\rho}^{\alpha}=\dot{\rho}^{\alpha}_{\rm mult}+\dot{\rho}^{\alpha}_{\rm ann}
\label{rot}
\end{equation}
The evolution model is initialized by the dislocation density at $t=0$, $\rho_0^{\alpha}$.
In eq.\ \eqref{rot}, $\dot{\rho}^{\alpha}_{\rm mult}$ and $\dot{\rho}^{\alpha}_{\rm ann}$ represent the dislocation multiplication and dislocation annihilation rate terms, respectively. In this model, both $\dot{\rho}^{\alpha}_{\rm mult}$ and $\dot{\rho}^{\alpha}_{\rm ann}$  are directly proportional to the plastic strain rate. Dislocation multiplication is treated as being proportional to the inverse mean free path of the dislocations, $\lambda_{\alpha}$:
\begin{equation}
\dot{\rho}^{\alpha}_{\rm mult}=\frac{|\dot{\gamma}^{\alpha}|}{b\lambda^{\alpha}}
\label{lambda}
\end{equation}
which is defined as a function of the grain size $d_g$, the \emph{forest} dislocation density $\rho_f^{\alpha}$, and a hardening constant $c$:
\begin{equation}
\frac{1}{\lambda^{\alpha}}=\frac{1}{d_g}+\frac{\sqrt{\rho_f^{\alpha}}}{c}
\label{mult_dislo}
\end{equation}
Here, $c$ and $d_g$ are set, respectively, to one and to an arbitrarily high value such that the term controlling the dislocation mean free path is:
$$\lambda^{\alpha}\approx\left(\sqrt{\rho_f^{\alpha}}\right)^{-1}$$
The forest dislocation density is calculated as \citep{roters2010overview}: 
\begin{equation}
\rho_f^{\alpha}=\sum_{\beta}\rho^{\beta}|\vec{n}^{\alpha}\cdot\vec{m}^{\beta}|
\end{equation}
Note that, in general, the mean free path as defined in eq.\ \eqref{mult_dislo} need not be equal to the effective dislocation segment length (in fact, it can be up to several orders of magnitude different \citep{basinski1979}). However, our model is designed with well-annealed, high-purity single W crystals in mind, with low initial dislocation densities and no impurities or obstacles other than dislocations themselves. Under this assumption, the use eq.\ \eqref{mult_dislo} can be justified in this case \citep{stainier_bcc_JMPS2002,Monnet2011451}.

For its part, dislocation annihilation occurs spontaneously when dipoles approach to within a spacing of $d_{\rm edge}$:
\begin{equation}
\dot{\rho}^{\alpha}_{\rm ann}=-\frac{2d_{\rm edge}}{b}\rho^{\alpha}|\dot{\gamma}^{\alpha}|
\label {ann}
\end{equation}

Equations \eqref{rot} through \eqref{ann} form the basis of the \emph{Kocks-Mecking} family of dislocation density evolution models \citep{mecking1981}. These models offer two interesting connections with the broader CP formulation employed here. First, a relation between the dislocation density evolution model and Section \ref{mobb} is established by way of the dislocation mean free path $\lambda^{\alpha}$, which determines the available segment length in the dislocation mobility function (eq.\ \eqref{mob1}). In this fashion, the dislocation velocity --and, through it, the plastic strain rate--  is self-consistently linked to the microstructure changes predicted by the model. Second, by virtue of the existence of latent and self-hardening, the model provides a correction to the available RSS for dislocation motion in eq.\ \eqref{mob1} of the following form:
\begin{equation}
\tau'^{\alpha}=\tau^{\alpha}-\tau_h=\matr{P}_{\rm tot}^{\alpha}:\matr{\sigma}-\mu b\sqrt{\sum_{\alpha'}\xi_{\alpha\alpha'}\rho^{\alpha'}}
\label{prime}
\end{equation}
where $\tau_h$ is the hardening stress and $\xi_{\alpha\alpha'}$ are the coefficients of the interaction matrix, which characterizes the interaction strength between slip systems $\alpha$ and $\alpha$' as a result of six possible independent interactions \citep{franciosi1983,franciosi1985}: self, coplanar, collinear, mixed-asymmetrical junction (orthogonal), mixed-symmetrical junction (glissile) and edge junction (sessile) \citep{madec_kubin_2004}.
The values of $\xi_{\alpha\alpha'}$ employed here are given in Table \ref{table1}, and were obtained from dislocation dynamics simulations of isotropic elastic bcc Fe under uniaxial deformation\footnote{Although the $\xi_{\alpha\alpha'}$ coefficients were calculated for bcc Fe and not W, the results are equally applicable because Fe was treated as isotropic elastic --as is W-- and the interaction matrix coefficients are non-dimensional and independent of the value of the plastic constants considered.} \citep{Queyreau2009}. The correspondence between each coefficient and each slip system considered in this work is given in \ref{app:slip}. 
\begin{table}[h]
\caption{List of parameters and functional dependences for fitting the CP model. All of these parameters have been obtained using dedicated atomistic calculations. The parameter $s$ represents the normalized shear stress: $s=\frac{\tau'^{\alpha}}{\sigma_P}$ (cf.\ eq.\ \eqref{prime}).}
\centering
\begin{tabular}{llc}
\hline
{\small parameter} & {\small value or function} & {\small units} \\
\hline\vspace{-3mm}\\
$a_0$ & 3.143 & \AA\\
$b$ & 2.72 & \AA\\
$h$ & $a_0\sqrt{6}/3$ & \AA\\
${\mathbb C}_{11}$ & 523 & GPa \\
${\mathbb C}_{12}$ & 202 & GPa \\
${\mathbb C}_{44}$ & 161 & GPa \\
$\nu_0$ & $9.1\times10^{11}$ & s$^{-1}$ \\
$\sigma_P$ & 2.03 & GPa \\
$B$ & $8.3\times10^{-5}$ & Pa$\cdot$s \\
$\Delta H(s;T)$ & $\Delta H_0\left(1-s^p\right)^q$ & eV \\
$\Delta H_0$ & 1.63 & eV \\ 
$p$ & 0.86 & - \\
$q$ & 1.69 & - \\
$w$ & 11 & $b$ \\
$\sigma_c^\chi$ & $\frac{\tau^{\ast}_c+\sigma\left(a_2\sin(2\chi)+a_3\sin(2\chi+\pi/6)\right)}{\cos\chi+a_1\cos\left(\pi/3+\chi\right)}$ & GPa 
\vspace{2mm}\\
$a_1$ & 0.938 & - \\
$a_2$ & 0.71 & - \\
$a_3$ & 4.43 & - \\
$\tau^{\ast}_c$ & 2.92 & GPa \\
$c$ & 1 & - \\
$d_g$ & 2.72 & \AA\\
$d_{\rm edge}$ & 2.72 & \AA\\
\hline
\end{tabular}
\label{table2}
\end{table}
\begin{table}[h]
\caption{Values of $\xi_{\alpha\alpha'}$ for latent hardening in bcc crystals (from \citet{Queyreau2009}).}
\begin{center}
\begin{tabular}{|cccccc|}
\hline
self & coplanar & collinear & orthogonal & glissile & sessile\\ 
\hline
0.009 & 0.009 & 0.72 & 0.05 & 0.09 & 0.06 \\
\hline
\end{tabular}
\end{center}
\label{table1}
\end{table}%
$\tau'^{\alpha}$ replaces $\tau^{\alpha}$ in eqs.\ \eqref{monkey1} to \eqref{mob1}, although, as mentioned earlier, this pertains mainly to the plastic flow regime and --as such-- is not expected to have a significant bearing on our calculations of $\sigma_y$.

\section{Results}
\label{section_results}

In this Section we present results of uniaxial and biaxial tensile test simulations to explore the dependence of the yield strength on loading direction, temperature and strain rate. First, however, a robust and consistent yield criterion must be defined to extract the yield stress from the raw output data from DAMASK.

\subsection{Yield criterion}
\label{sec:yy}

In metals, where dislocation flow is not a singular event but a diffuse continuous process, it is generally accepted that the definition of yield point\footnote{Also referred to as elastic limit, proportionality limit, yield stress, etc.} is not unique. Perhaps as the result of these conceptual indetermination, modern usage has evolved into that of an arbitrary rule, the 0.2\% strain offset rule for obtaining the yield stress of metals. For materials having nonlinear elastic behavior, there are not even arbitrary rules, only individual preferences and proclivities in defining yield when a given amount of strain has been reached.
It is quite apparent then that to define robust yield criteria it is necessary that they be implemented and supported by consistent and meaningful definitions in terms of the stress-strain behavior. This is often difficult when the transition from the elastic to the inelastic regimes is obscured in the global picture of deformation. However, in the present calculations we effectively possess an arbitrary degree of data resolution and can define an unambiguous mathematical criterion.  

The preferred method for defining the elastic limit of a ductile material is to compute the second derivative of the stress-strain curve, referred to generically as $\sigma(\varepsilon)$, and identify the location of the inflection point \citep{christensen2008}. The yield point then corresponds to the strain, $\varepsilon_y$, for which $\left|\frac{d^2\sigma}{d\varepsilon^2}\right|$ is maximum. Mathematically:
\begin{equation}
\sigma_y=\sigma(\varepsilon_y),~\varepsilon_y:=\varepsilon\mid \max\left|\frac{d^2\sigma}{d\varepsilon^2}\right|
\label{eq:crit}
\end{equation}
For ductile metals, the location of the maximum of the second derivative represents the point at which dislocation-mediated flow is the major contribution to $\matr{L}$ (cf.\ Section \ref{flowsec}). However, this condition works surprisingly well for other materials such as glassy polymers, where flow might be caused by molecular rearrangement and damage at both the molecular and macroscopic scales \citep{bowden1972}.

\begin{figure}[h]
\centering
\includegraphics[width=0.9\textwidth]{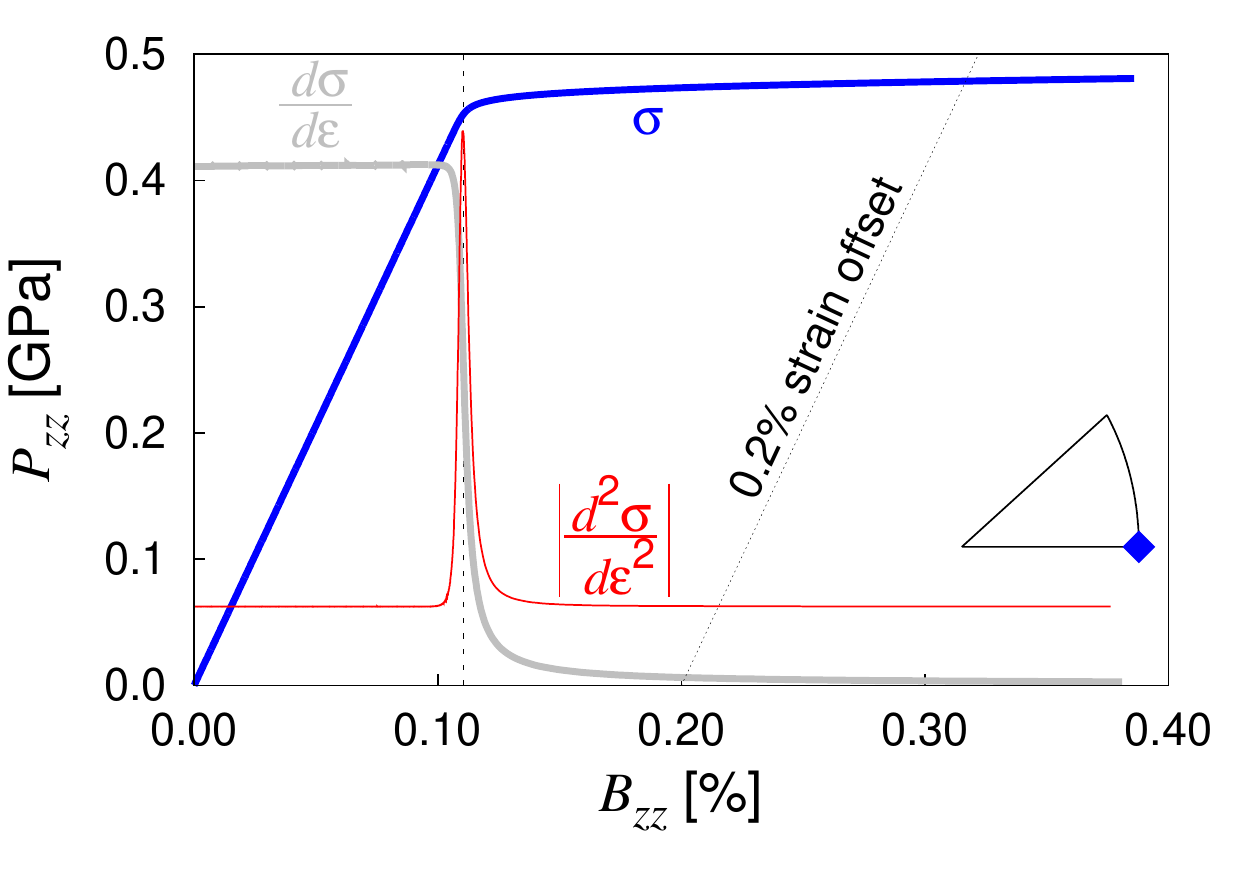}
\caption{\label{fig_P1_vs_Biot} Evolution of the stress $P_{zz}$ with deformation $B_{zz}$ during a CP simulation of a uniaxial tensile test with $[101]$ loading orientation (as depicted in the standard triangle). The first and second derivatives of the stress w.r.t.~to the strain are also plotted to illustrate the method of identification of the yield point according to this criterion. Also shown is the intercept of the curve with the 0.2\% strain offset criterion line.} 
\end{figure}

To illustrate the accuracy of the second-derivative method, we plot in Figure \ref{fig_P1_vs_Biot} the first and second derivative of a stress-strain curve corresponding to a $[101]$ uniaxial tensile test of a W single crystal under representative initial conditions. Recall from Section \ref{CP_constitutive_model} that the stress and strain metrics of choice are $\matr{P}$ and $\matr{B}$, and so we plot $\frac{dP_{zz}}{dB_{zz}}$ and $\frac{d^2P_{zz}}{dB_{zz}^2}$ specifically.
The inflection point --marked by a vertical dashed line in the figure-- occurs for $\varepsilon_y=0.1105\%$, for which a value of $\sigma_y=0.452$ GPa is obtained. The figure also shows the 0.2\% strain offset criterion, which --by contrast-- gives $\varepsilon_y=0.3167\%$ and $\sigma_y=0.479$ GPa, {\it i.e.}~a three-fold difference in strain and approximately a 6\% difference in stress with respect to the stress second derivative criterion.

However, determining the first and second derivatives of the stress-strain relation can become numerically intensive, especially when evaluating thousands of curves as is the case in this work. An approximation to this method that works particularly well for linear-elastic materials that display a clear elastic-to-plastic transition is to take the yield point as the first point in the $\sigma({\varepsilon})$ function that satisfies:
$$\frac{d\sigma}{d\varepsilon}<E(1-\delta)$$
{\it i.e.}~$\sigma_y$ is measured as the stress for which a departure from linearity (as set by the elastic regime) larger than some small value $\delta$ is observed in the stress-strain relation. We have found that a value of $\delta\approx0.01$ is sufficient to predict the value of $\sigma_y$ within a small error relative to the value furnished by the second-derivative method. By way of example, for the curve shown in Fig.\ \ref{fig_P1_vs_Biot} and $\delta=0.01$, we find a values of $\varepsilon_y=0.1055\%$ and $\sigma_y=0.435$ GPa, or less than a 4\% difference with the numbers according to the second-derivative criterion. With this reasonable accuracy and the computational advantages alluded to above, we then use the $\delta=0.01$ criterion in the remainder of this paper.

\subsection{Model validation and initial results}
\label{validation}

\begin{figure}[h]
\centering
	\subfigure[\text{[111] loading}]{
			\includegraphics[width=0.7\linewidth]{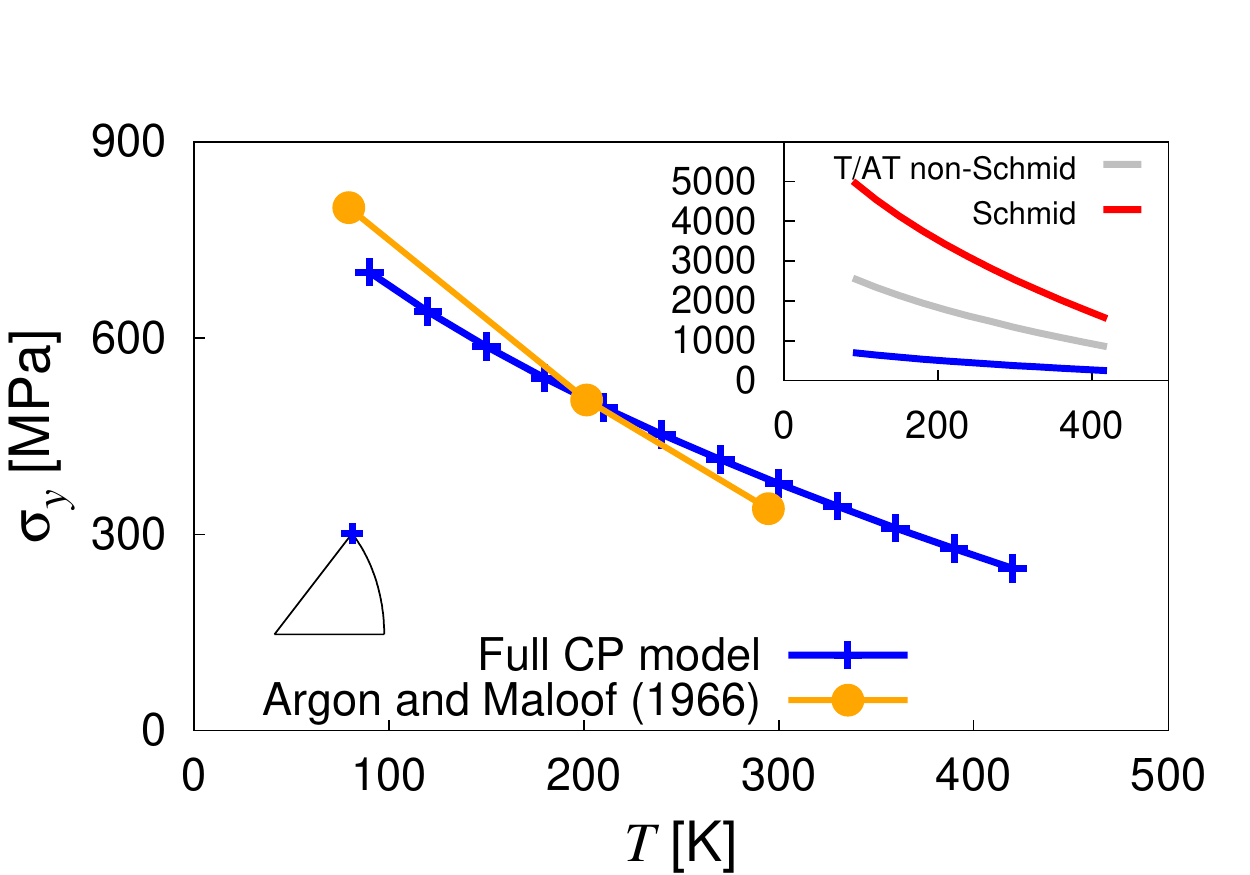}
    	\label{a1}}
    	\subfigure[\text{[110] loading}]{
        \includegraphics[width=0.7\linewidth]{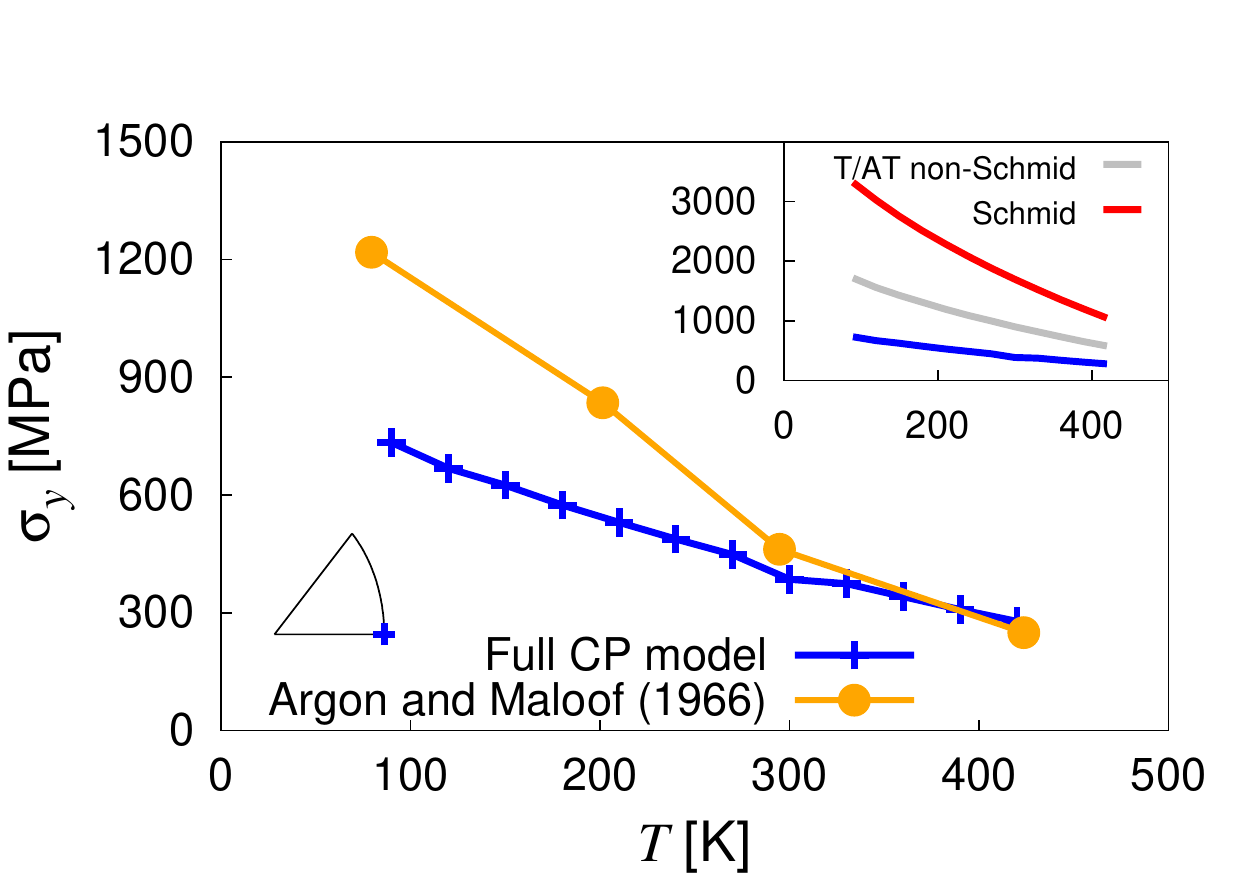}
    	\label{a2}}
	   \caption{\label{argon} Yield strength of W single crystals at the conditions used by \citet{argon_maloof_1966} in tensile deformation tests under two different loading orientations. The experimental data is shown for comparison. The inset shows the results of CP calculations with different contributions of the projection tensor activated.}
\end{figure}

Prior to deploying our fully-parameterized CP method for numerically-intensive calculations, it is essential to undergo a thorough exercise of validation. Experimental data from uniaxial tensile tests in single crystal W at low strain rates are scant and sporadic, with the main sources listed below:
\begin{enumerate}
\item \citet{argon_maloof_1966} performed some early experiments at a strain rate of $10^{-4}$ s$^{-1}$ and temperatures of 77, 199, 293, 373, and 450 K. These authors measured the yield strength for the three vertices of the stereographic triangle $[001]$, $[110]$, and $[111]$ with an initial dislocation density of $\rho_0\approx10^{10}$ m$^{-2}$.
\item \citet{raffo1969} analyzed the yielding behavior of arc-melted W between 77 and 680 K at $\dot{\varepsilon}=8.3\times10^{-4}$ s$^{-1}$. However, the loading orientation is not given and most of the tests were done in compression. 
\item \citet{stephens1970} has carried out compression tests at 150, 300, and 590 K. This researcher focuses on dislocation density evolution and dislocation substructures, however, with a value of $\rho_0\approx1.4\times10^{14}$ m$^{-2}$, notably larger than in other tests. There have been other works that have also focused mainly on compression tests \citep{raffo1965,gupta1970}. 
\item \citet{Brunner,Brunner_Wtemperature2010} has performed a series of experiments more recently at temperatures between 77 and 800 K. They employed a value of $\dot{\varepsilon}=8.5\times10^{-4}$ s$^{-1}$  and loaded the system uniaxially along the $[\bar{1}4~9]$ direction with a starting dislocation density of $5.5\times10^9$ m$^{-2}$.  
\end{enumerate}
As pointed out in Section \ref{sec:schmid}, our CP model is parameterized for tensile tests only and so for validation we focus on the works by \citet{argon_maloof_1966} and \citet{Brunner,Brunner_Wtemperature2010}. \citet{argon_maloof_1966} centered on multislip by considering mainly loading orientations coincident with the vertices of the standard triangle. Consequently, we replicate their test conditions in our CP model and compare the results obtained by taking into account all the different elements of the projection tensor \eqref{pepe}. The results are shown in Figure \ref{argon} for the $[111]$ and the $[110]$ loading orientations, with the insets in both figures showing the relative importance of considering each of the non-Schmid contribution to the projection tensor incrementally. While our calculations are in general good agreement with the $[111]$ test data, they deviate from the experimental results at the two lower temperature points for the $[110]$ orientation. \citet{argon_maloof_1966} point out that, at low temperatures, deformation by twinning may play a larger role when loading along $[110]$ relative to other orientations. This may be at the origin of the discrepancy, as twinning is not part of the catalog of deformation mechanisms considered in this model.

\begin{figure}[h]
\centering
\includegraphics[width=1.0\linewidth]{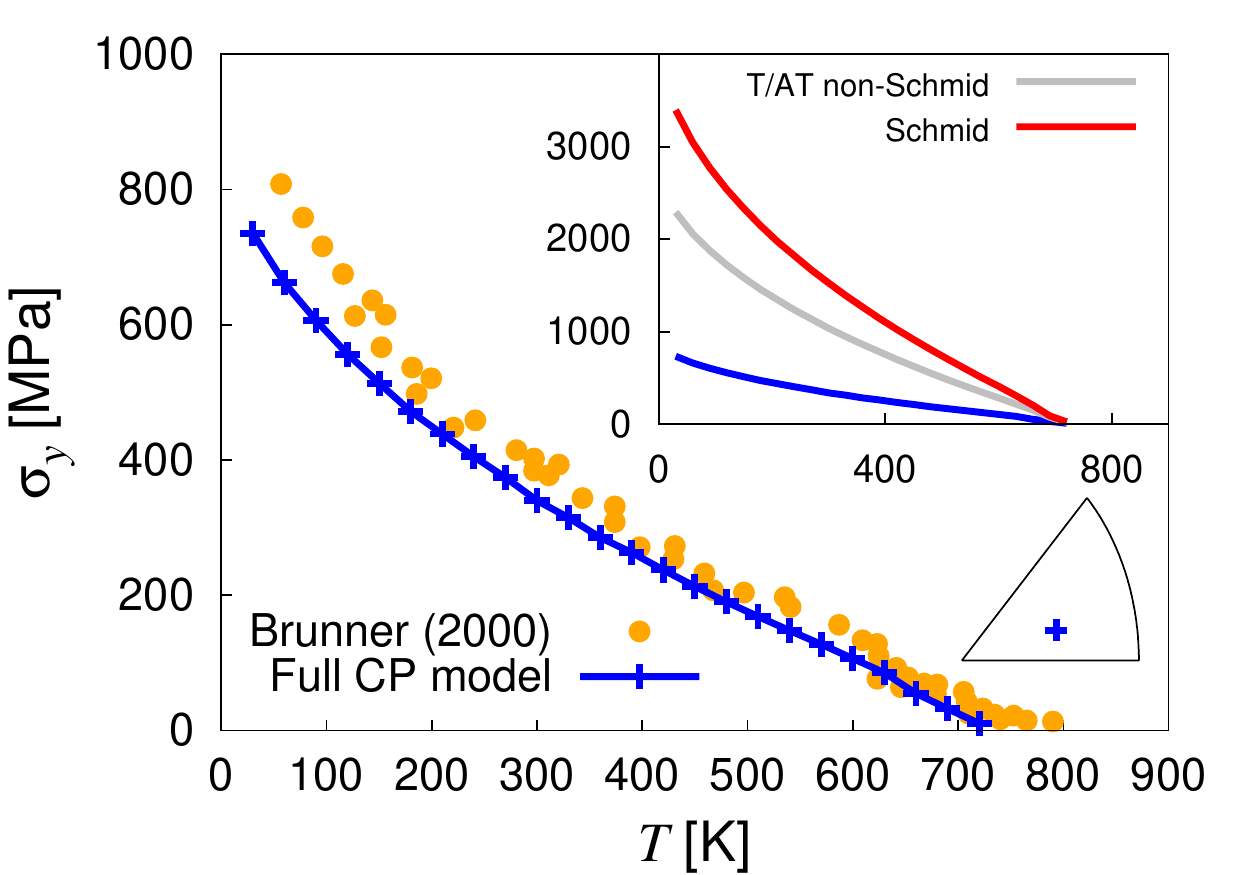}
\caption{\label{yield} Yield strength of W single crystals under the conditions used by \citet{Brunner_Wtemperature2010} in uniaxial tensile tests. The experimental data is shown for comparison. The inset shows the results of CP calculations with different contributions of the projection tensor activated.} 
\end{figure}

Next we simulate uniaxial tensile tests under single slip conditions, {\it i.e.} along crystal orientations near the center of the standard triangle. This corresponds to the experiments by \citet{Brunner,Brunner_Wtemperature2010} referred to above, which were done more recently with more advanced instrumentation. The results are shown in Figure \ref{yield}, where we also show the curves using the different elements of eq.\ \eqref{pepe}. This time, the agreement is striking, particularly again at temperatures above 400 K. Specifically, the \emph{athermal} limit ($\approx710$ K) is particularly well reproduced, as is the extrapolated critical stress at 0 K (Peierls stress), which is within 10\% of the experimental values.

Although, as noted earlier, the main focus of this work is on yielding, we have applied the fully parameterized model to study the flow stress regime for some selected cases in \ref{app_flow}. The results shown there demonstrate the performance of the method outside the primary range of application. While the model cannot be assumed to be predictive in the post-yield regime under general loading conditions, these are encouraging results that strengthen the notion that parameter-free CP calculations can perform well under specific deformation scenarios.

With the confidence conferred on our CP model by the validation procedure, next we proceed to calculate the yield strength for a number of numerically-intensive scenarios. This is the object of the following sections.

\subsection{Uniaxial tensile tests} 
\label{subsection_uniaxial}
In this Section, we report on the uniaxial yielding results as a function of temperature and strain rate. Our results are organized by strain rate, such that we first provide a detailed account of all the calculations at a given strain rate followed by a study on the dependence with $\dot\varepsilon$.   

\subsubsection{Results at $\dot\varepsilon=10^{-3}~\rm s^{-1}$ }

\begin{figure}[h]
        \centering
        \subfigure[100 K]{\includegraphics[width = 0.45\textwidth]{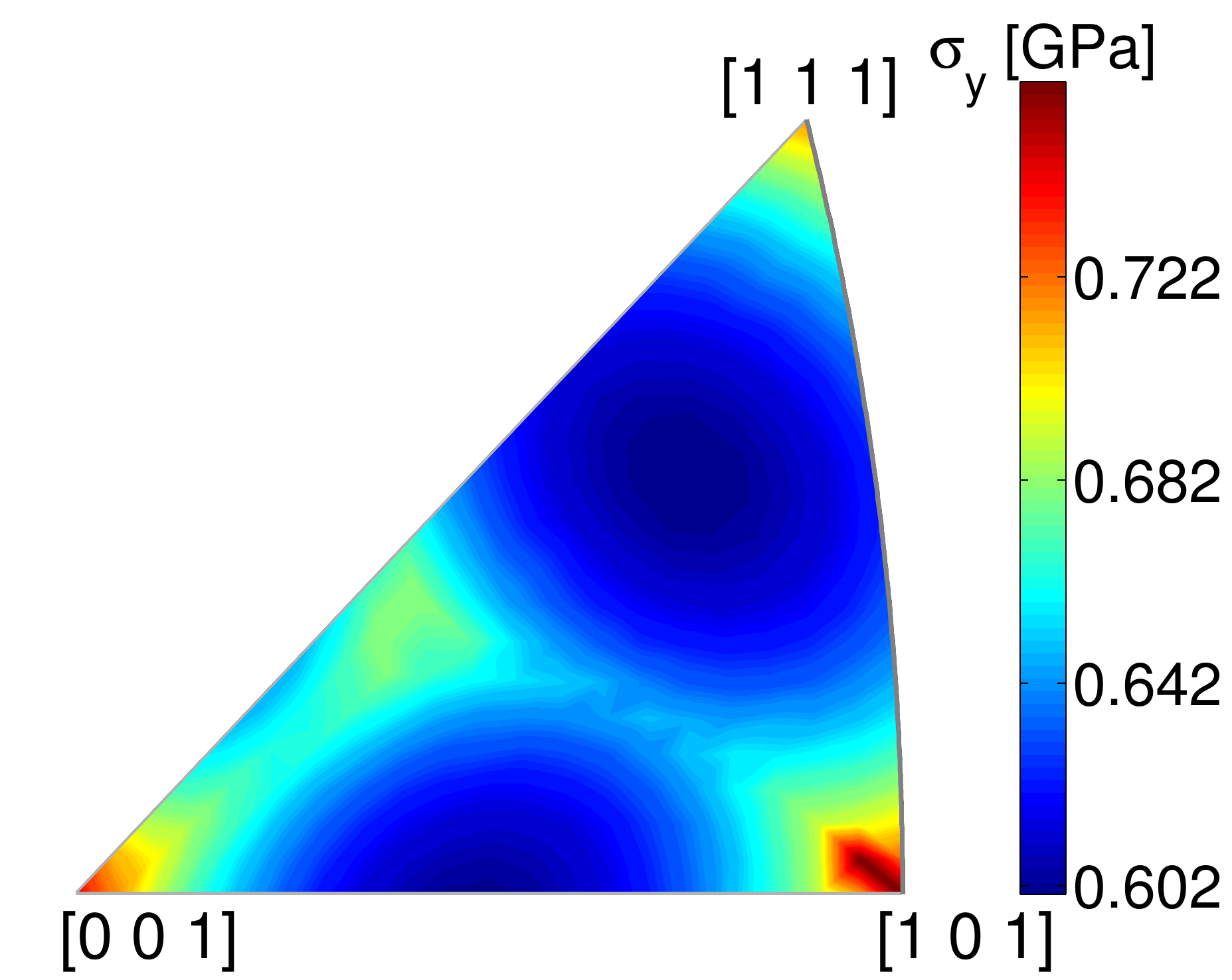}	\label{fig_N231_srate3_100} }
		\hfil
		\subfigure[200 K]{\includegraphics[width = 0.45\textwidth]{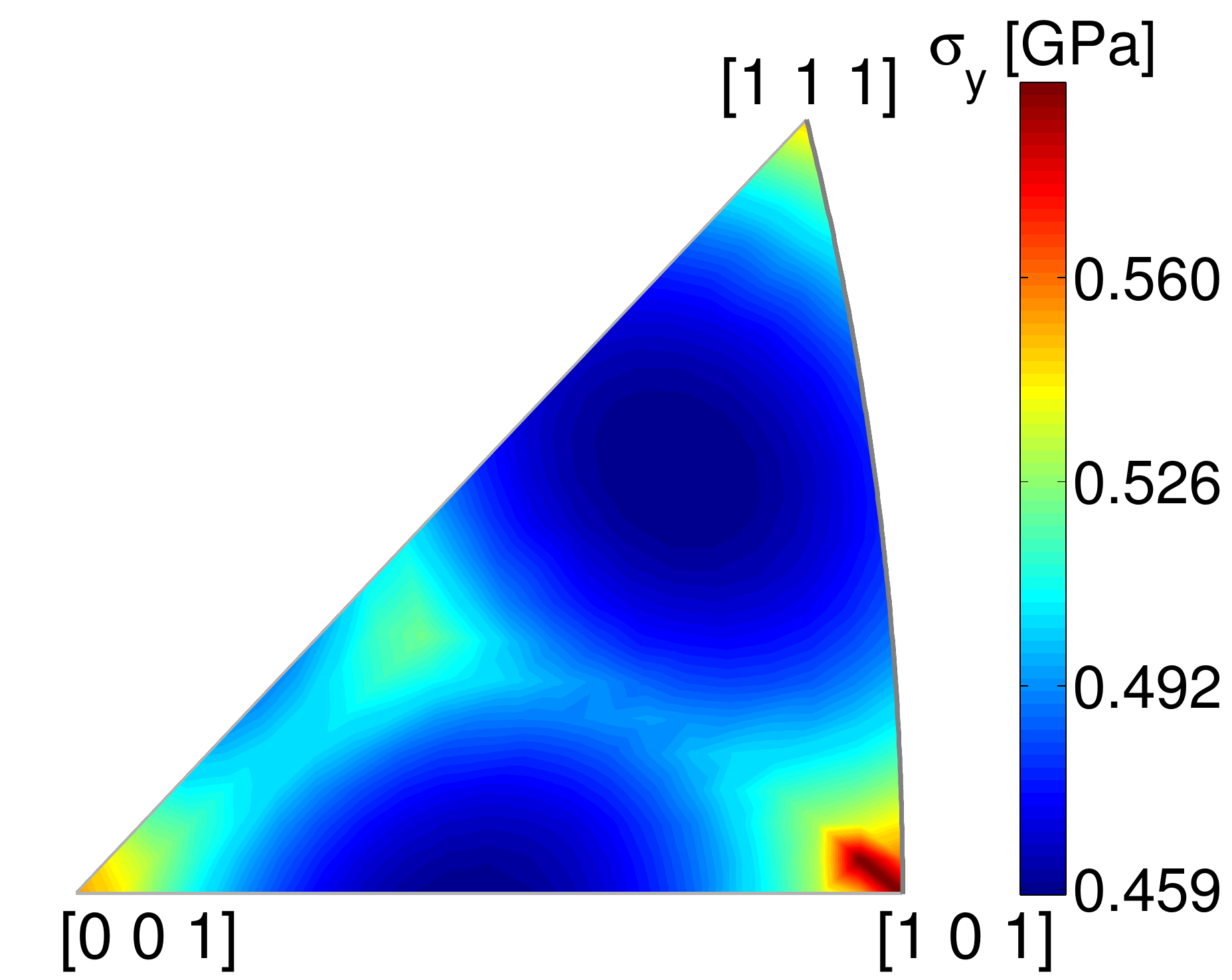}	\label{} }
		\hfil
		\subfigure[300 K]{\includegraphics[width = 0.45\textwidth]{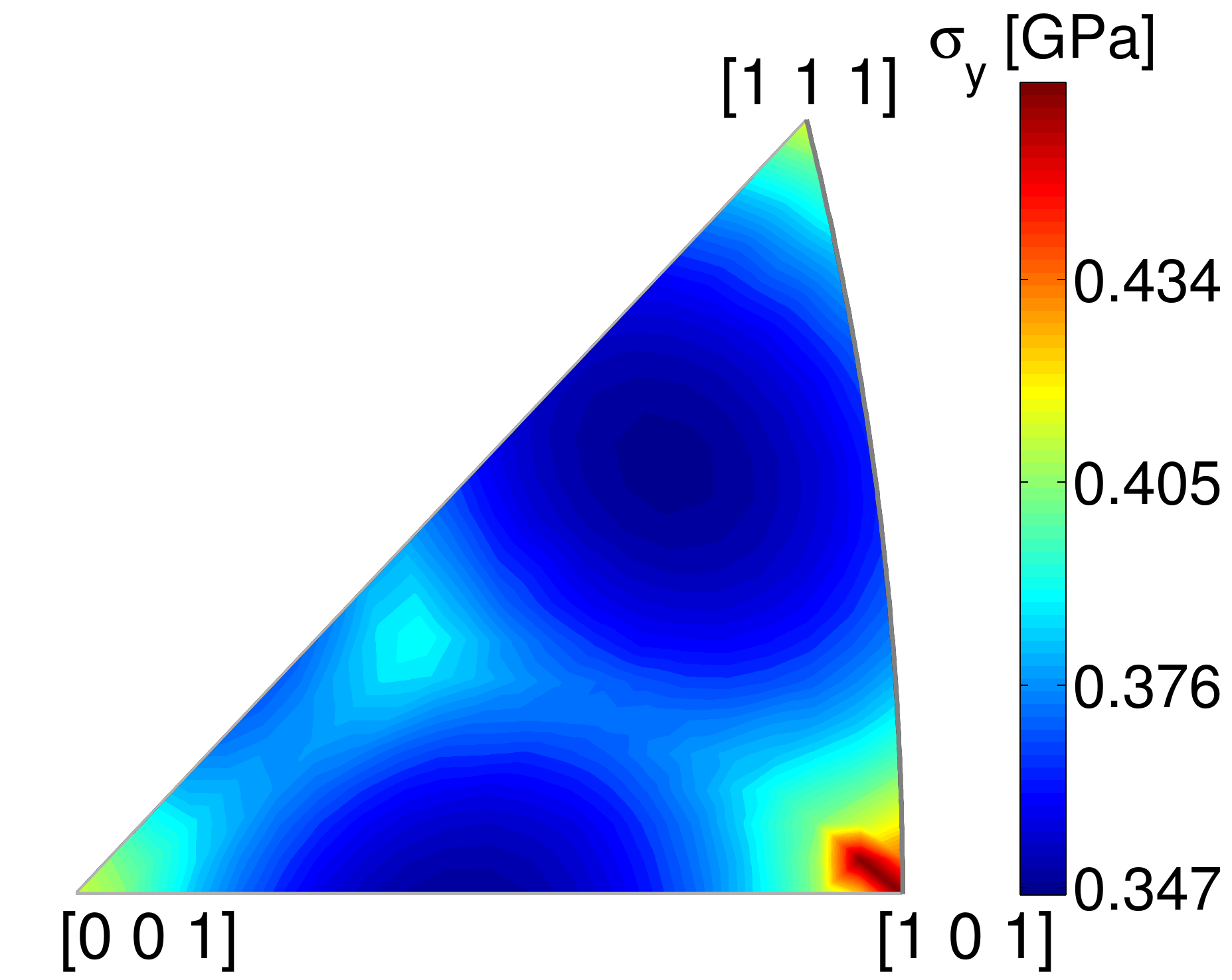}\label{} }   
		\hfil
		\subfigure[400 K]{\includegraphics[width = 0.45\textwidth]{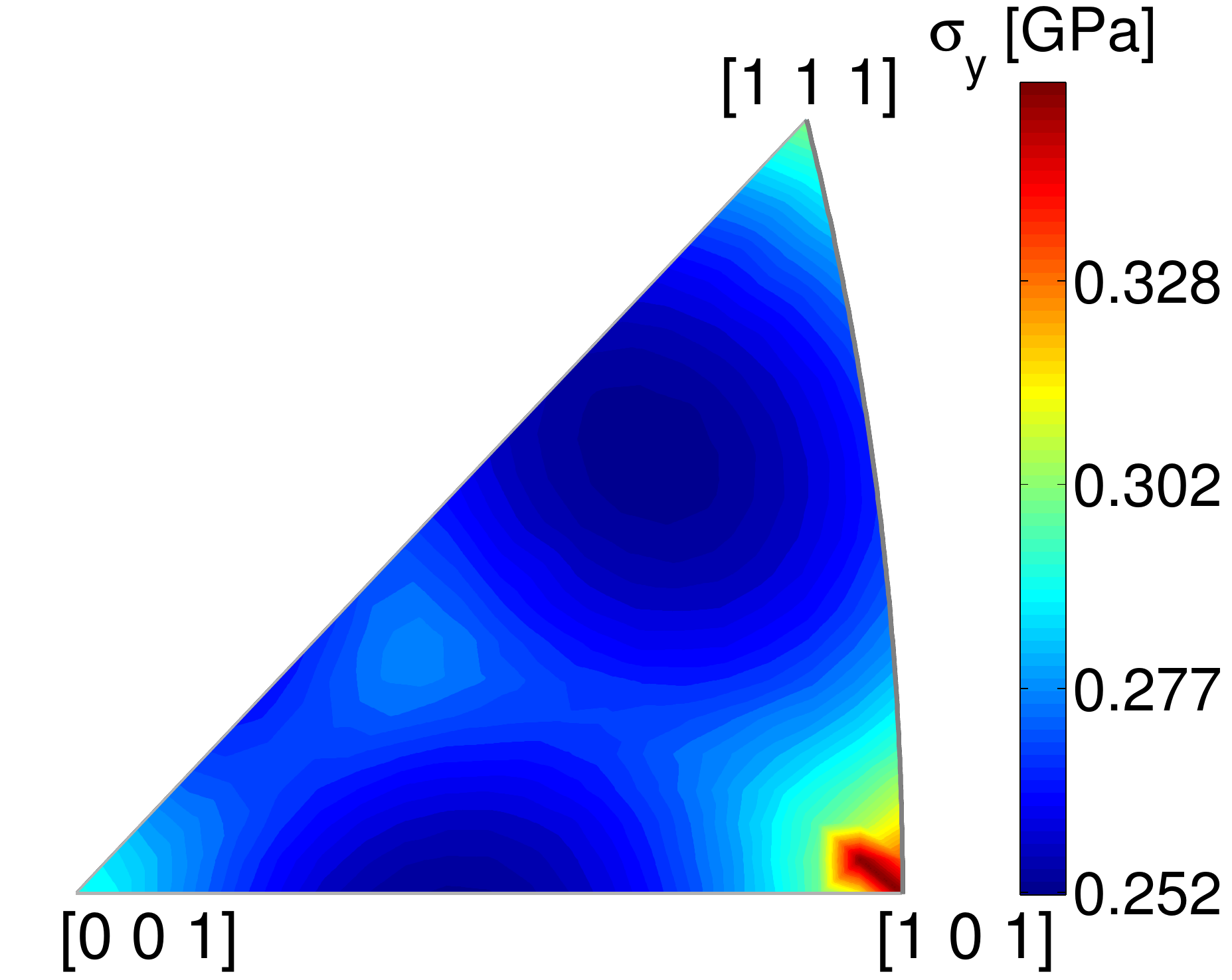}	\label{} }
		\hfil
		\subfigure[500 K]{\includegraphics[width = 0.45\textwidth]{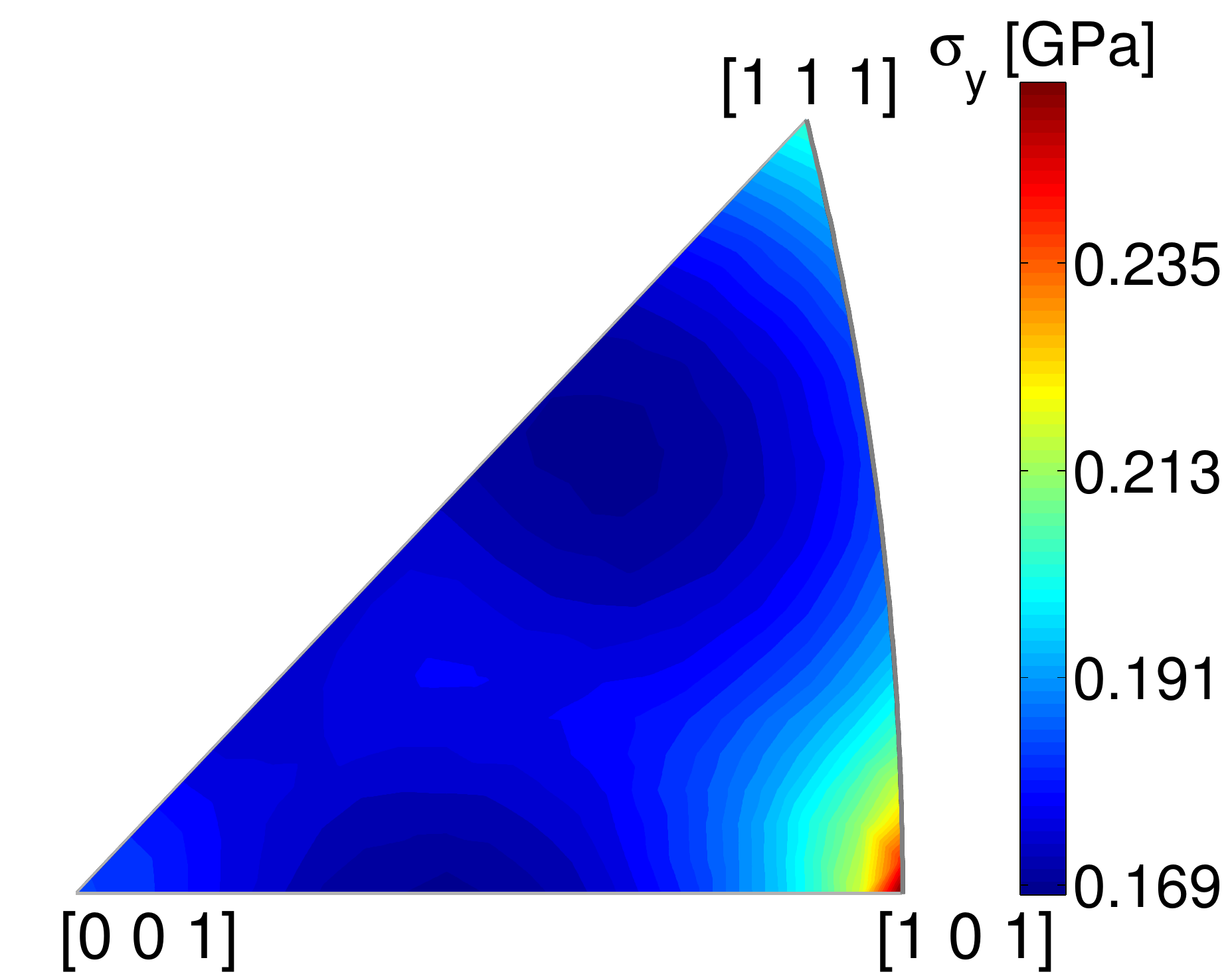}\label{} }   
		\hfil
		\subfigure[600 K]{\includegraphics[width = 0.45\textwidth]{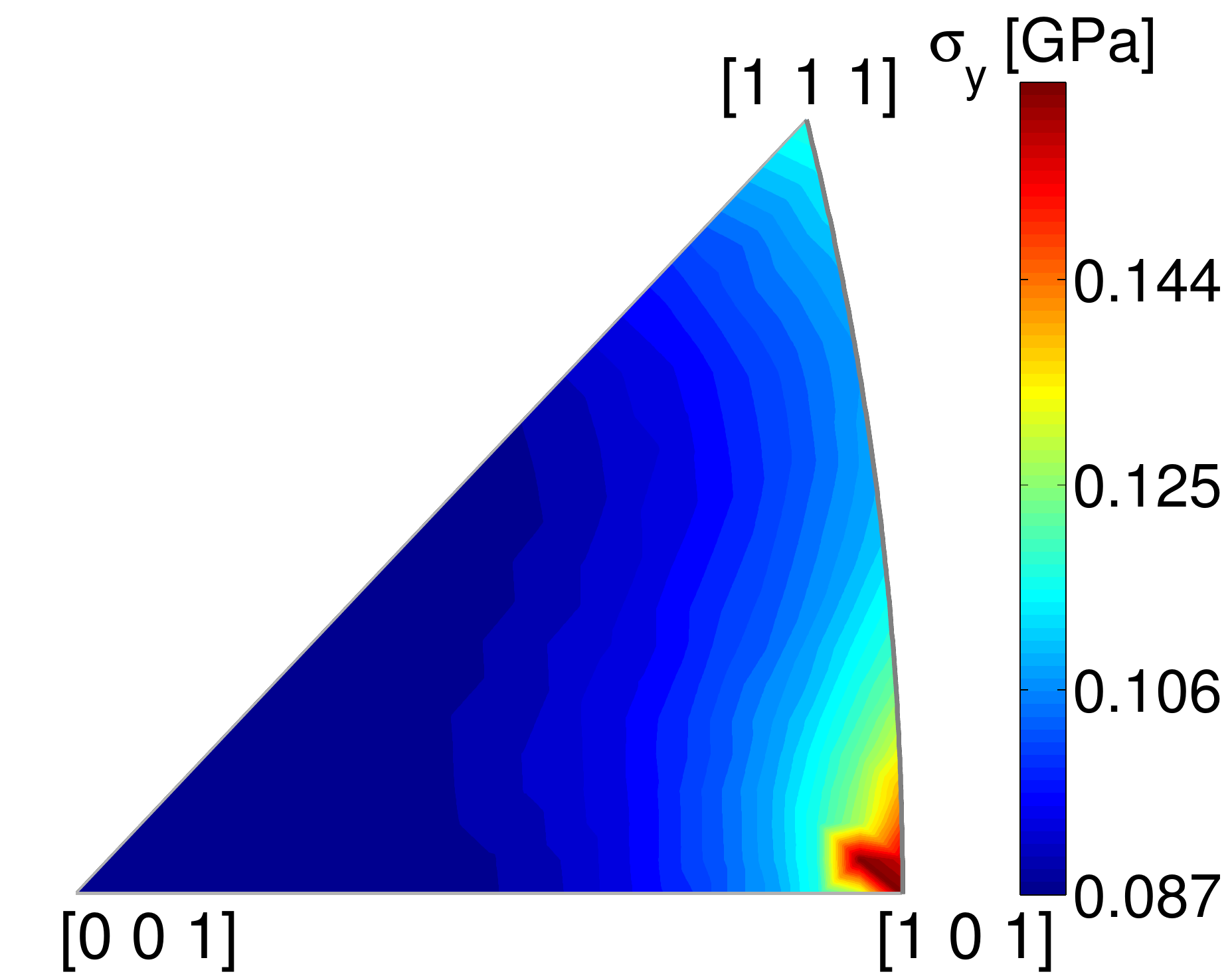} \label{fig_N231_srate3_600}	 }
        \caption{\label{fig_yield_strength_N231_srate3} Contour maps of the yield strength from uniaxial tensile test simulations for 231 uniformly distributed crystallographic orientations in the standard triangle at different temperatures. Note that each map has its own distinct numerical scale to aid in the visualization of hard and soft regions. }
\end{figure}

For these calculations, we have discretized the area of the standard triangle into a uniform grid consisting of 231 nodes, each representing a crystallographic loading orientation.
We begin with calculations at a prescribed strain rate of $\dot\varepsilon=10^{-3}~\rm s^{-1}$. Figure \ref{fig_yield_strength_N231_srate3} shows colored contour plots of the yield stress in the 100-to-600 K temperature range. Areas of high relative yield strength can be seen to concentrate around the vertices of the standard triangle, representing multislip conditions, whereas soft regions develop in two distinct locations of the triangle, one near the $[324]$ zonal axis that then rotates towards $[112]$ above 500 K, and another near $[102]$. Note that, to accentuate the differences between hard and soft regions, each contour plot has its own specific numerical scale. 

\begin{figure}[h]
        \centering
		\subfigure[$\dot\varepsilon=10^{-3}~\rm s^{-1}$]{\includegraphics[width = 0.3\textwidth]{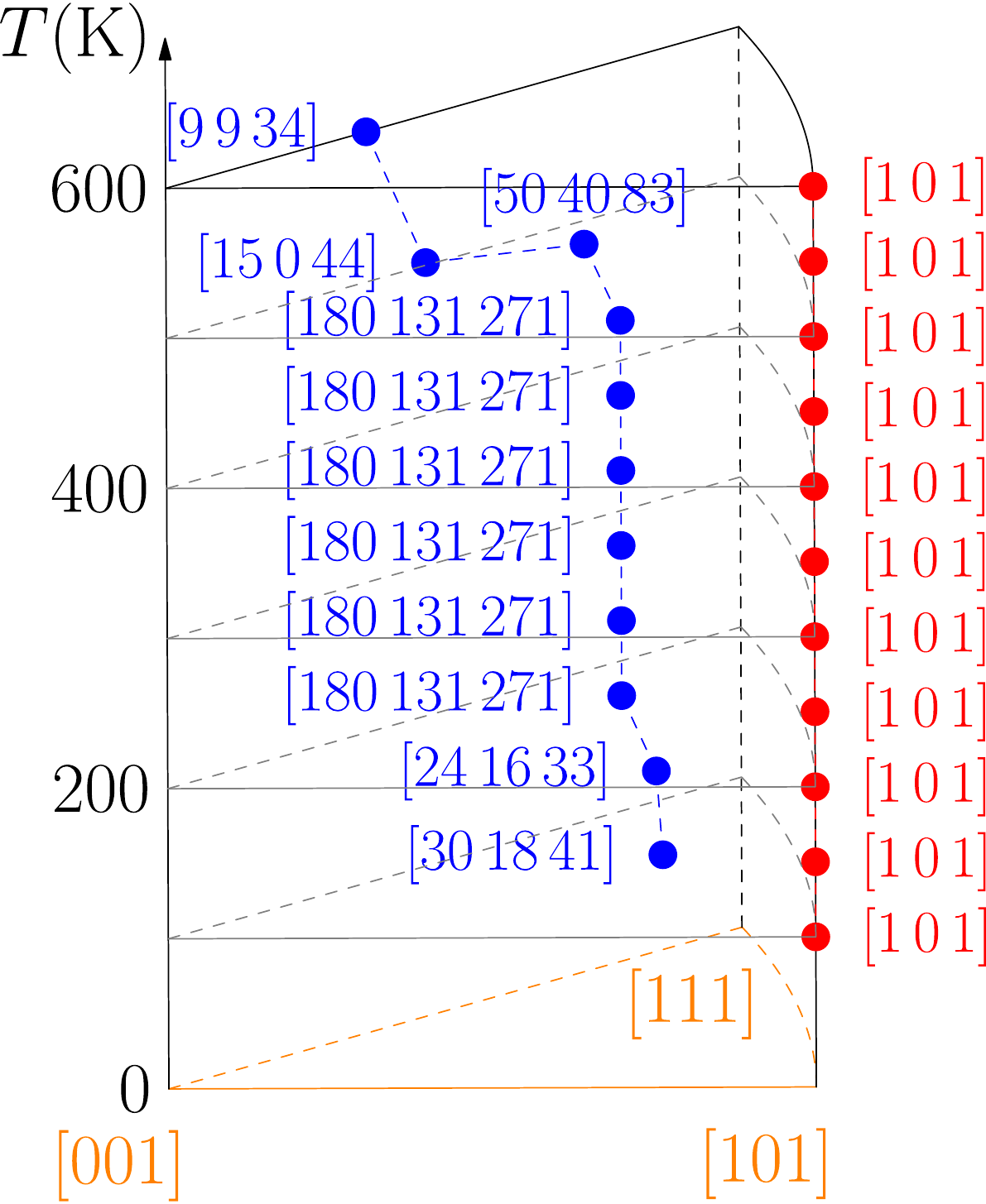}	\label{e3} }
		\hfil
		\subfigure[$\dot\varepsilon=10^{-4}~\rm s^{-1}$]{\includegraphics[width = 0.3\textwidth]{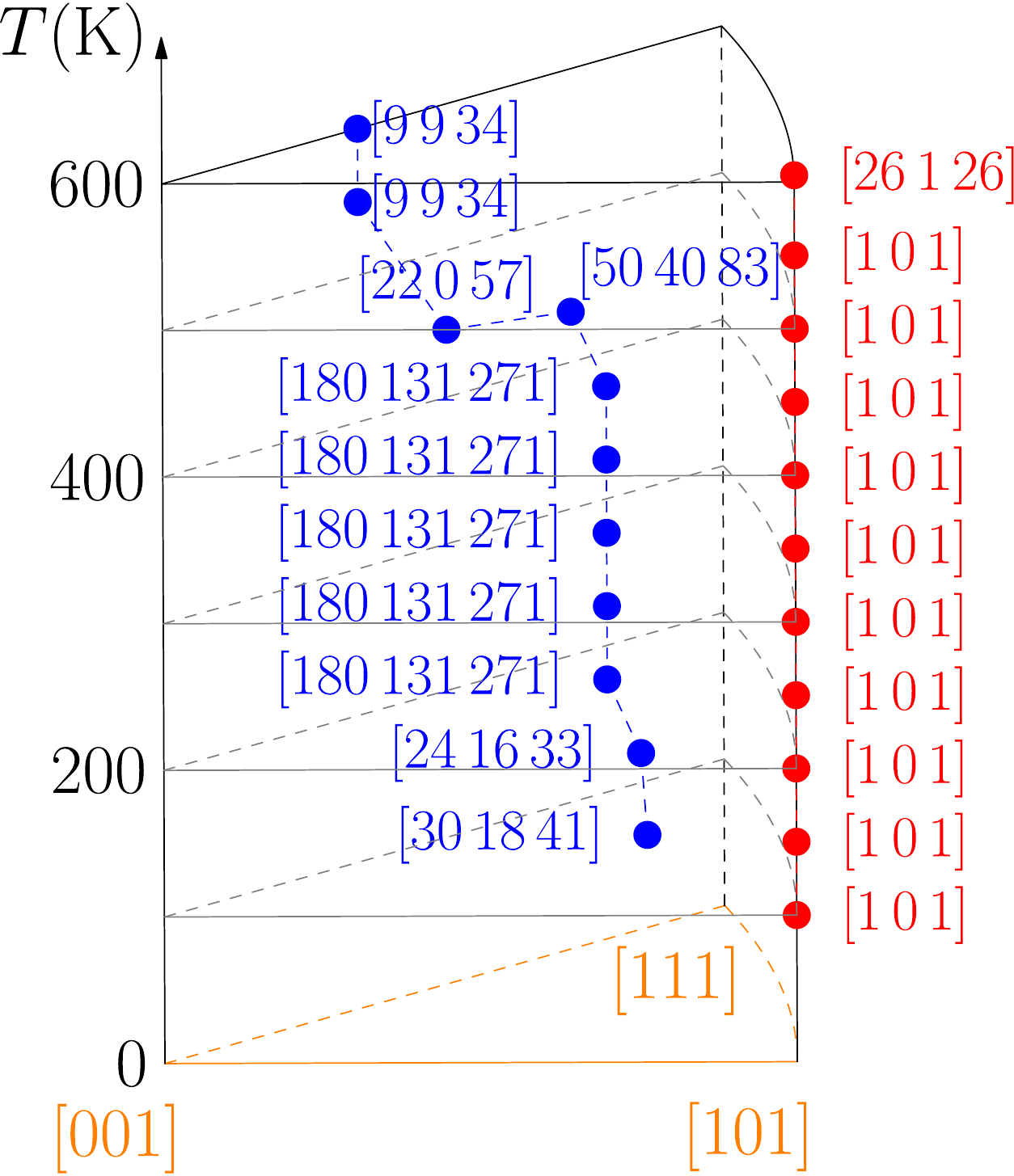}\label{e4} }  
		\hfil
		\subfigure[$\dot\varepsilon=10^{-5}~\rm s^{-1}$]{\includegraphics[width = 0.3\textwidth]{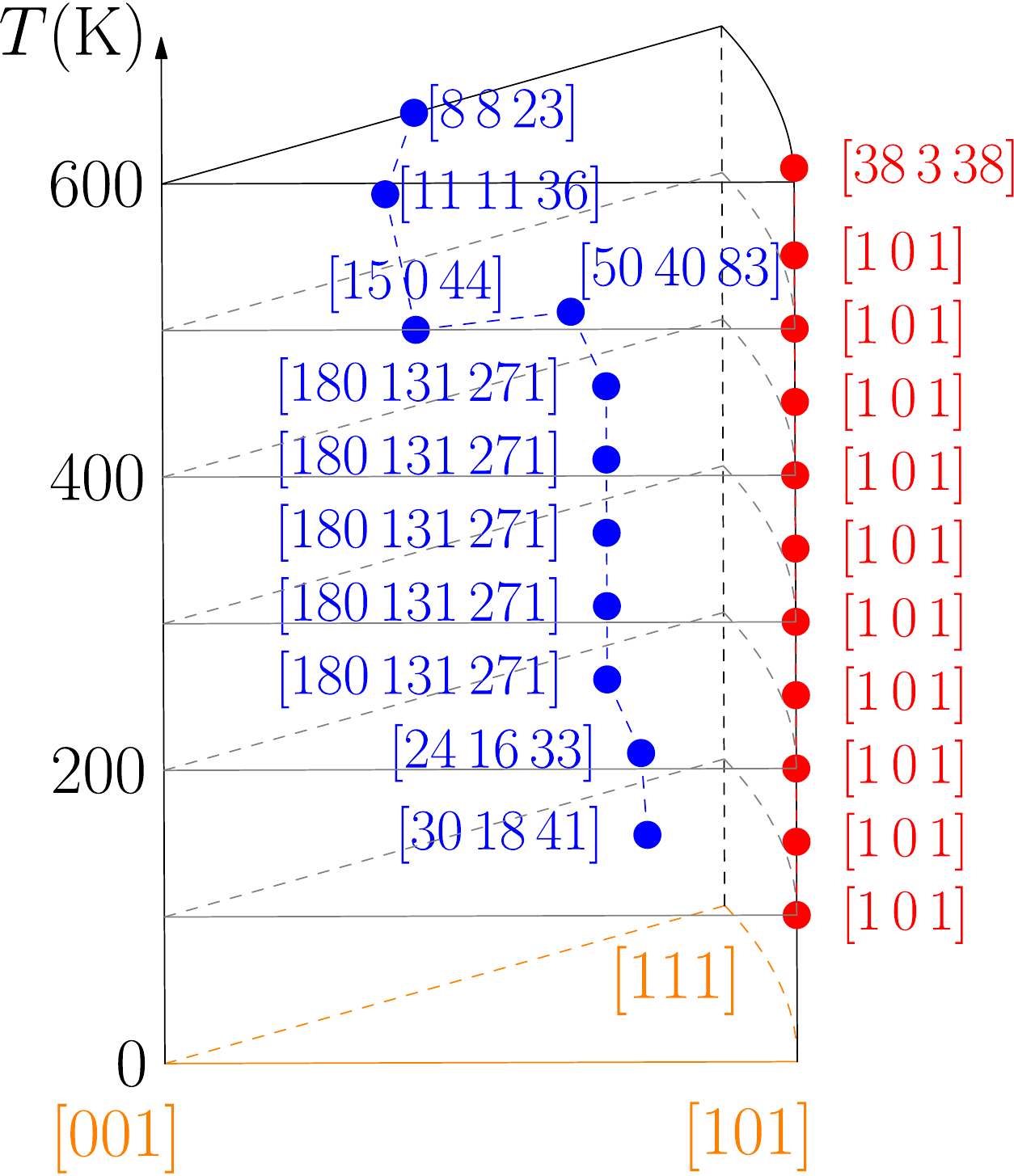}\label{e5} } 
        \caption{Temperature path of the softest and hardest yield directions on the standard triangle as a function of strain rate. \label{fig_max_min_loading_directions}}
\end{figure}

\begin{figure}[h]
        \centering
		\subfigure[]{\includegraphics[height = 0.38\textwidth]{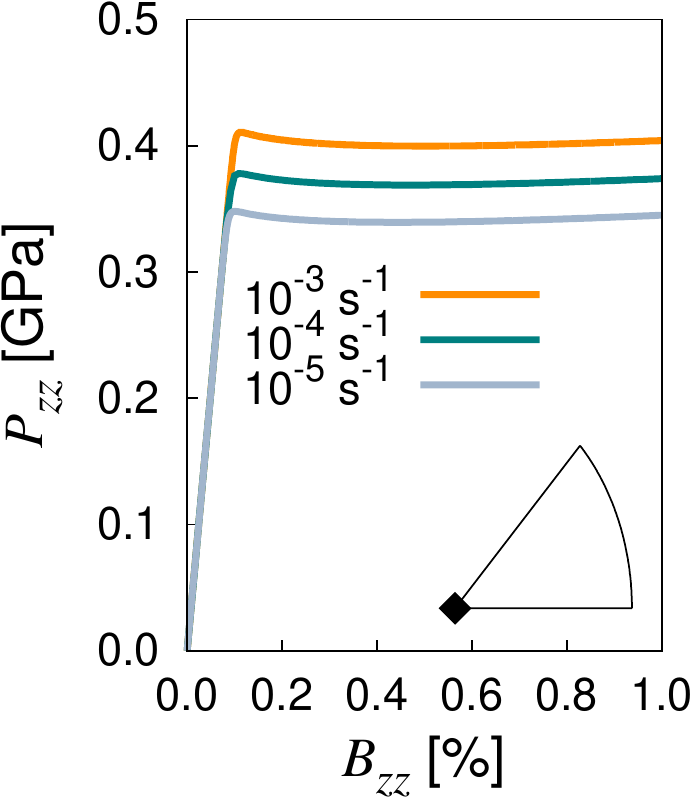}\label{s1} }
		\hfil
		\subfigure[]{\includegraphics[height = 0.4\textwidth]{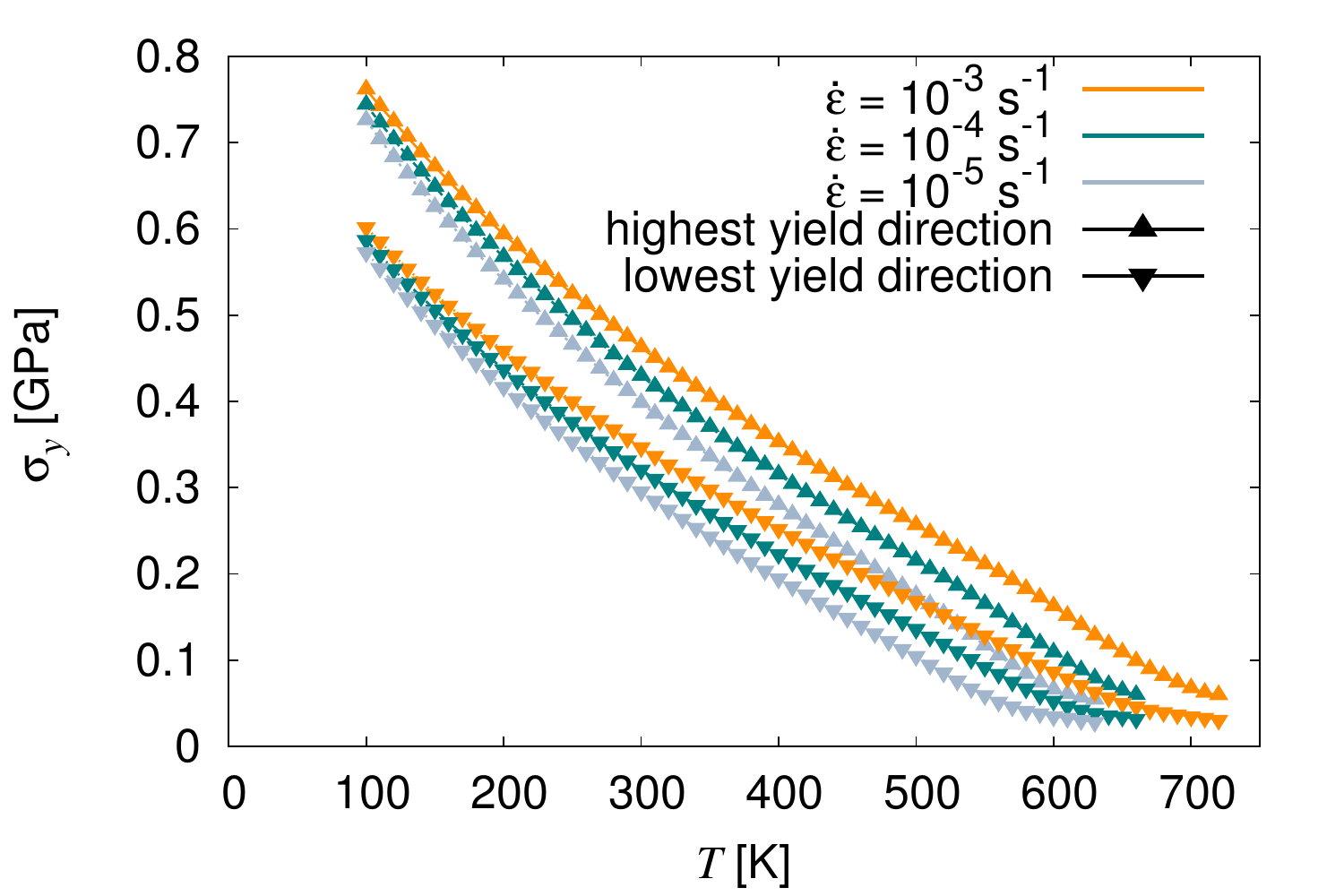}\label{s2} }  
        \caption{(a) Stress-strain relations at three different strain rates and $T=300$ K for a \hkl[001] loading orientation. (b) Temperature dependence of the yield strength for the softest and hardest directions as a function of strain rate. \label{fig_yield_vs_temp_vs_srate}}
\end{figure}

We have extracted the specific location of the global extrema in the standard triangle and plot it as a function of temperature in Figure \ref{e3}. The hardest direction is consistently the $[101]$, while the softest is seen to revolve around the vicinity of the $[112]$ axis, first along $[30~18~41]$ at 100 K, then along $[180~131~271]$ between 200 and 500 K, and finally rotating towards $[9~9~34]$ for $T>500$ K.
Next, we plot the detailed temperature dependence of the yield strength corresponding to the hardest and softest directions --as given by Fig.\ \ref{e3}-- for this strain rate in Figure \ref{fig_yield_vs_temp_vs_srate}. As the calculated data show, there is approximately a 30\% difference in yield stress between the hardest and softest directions. Interestingly, this gap appears to be fairly independent of temperature. Above 650 K, the curves begin to level off, signaling the onset of the athermal regime.

\subsubsection{Dependence on strain rate and strain rate sensitivity}
\label{mmmm}
In this Section we expand the analysis presented in the previous Section to strain rates of $10^{-4}$ and $10^{-5}~\rm s^{-1}$. To avoid redundancies, here we show only the temperature trajectory of the softest and hardest loading orientations in Figs.\ \ref{e4} and \ref{e5}, which emanate from calculations as those presented in Fig.\ \ref{fig_yield_strength_N231_srate3}. The results are quantitative similar to the case of $\dot\varepsilon=10^{-3}$ s$^{-1}$, with the only appreciable deviations occurring at temperatures above 450 K. At these high temperatures, the softest orientation rotates clearly towards the vicinity of the $[113]$ zonal axis, without excursions near $[103]$ as was the case for the  $\dot\varepsilon=10^{-3}$ calculations. 

As above, we add the temperature dependence of the yield stress for the hardest and softest directions at these strain rates to Figure \ref{fig_yield_vs_temp_vs_srate}. The data show the same qualitative trend for all strain rates, with the same approximate 30\% difference between the hard and soft orientations. 
However, useful information can be extracted if the strain-rate dependence of the yield stress is plotted for selected orientations. Then, one can calculate the so-called \emph{strain rate sensitivity}, characterized by the strain rate sensitivity exponent $m$, of the material as a function of temperature. Strain rate sensitivity is exceedingly important to delay the onset of inhomogeneous deformation \citep{hutchinson1977}, {\it e.g.}~necking, and is used as a criterion to assess the possibility of superplastic behavior in certain kinds of materials \citep{hedworth1971,arieli1976}. This belongs more in the realm of failure and is thus outside the scope of this paper. However, it is of interest to calculate the strain rate sensitivity of the yield stress and relate our findings to the larger failure picture if possible.

\begin{figure}[h]
        \centering
        \includegraphics[width = 0.7\textwidth]{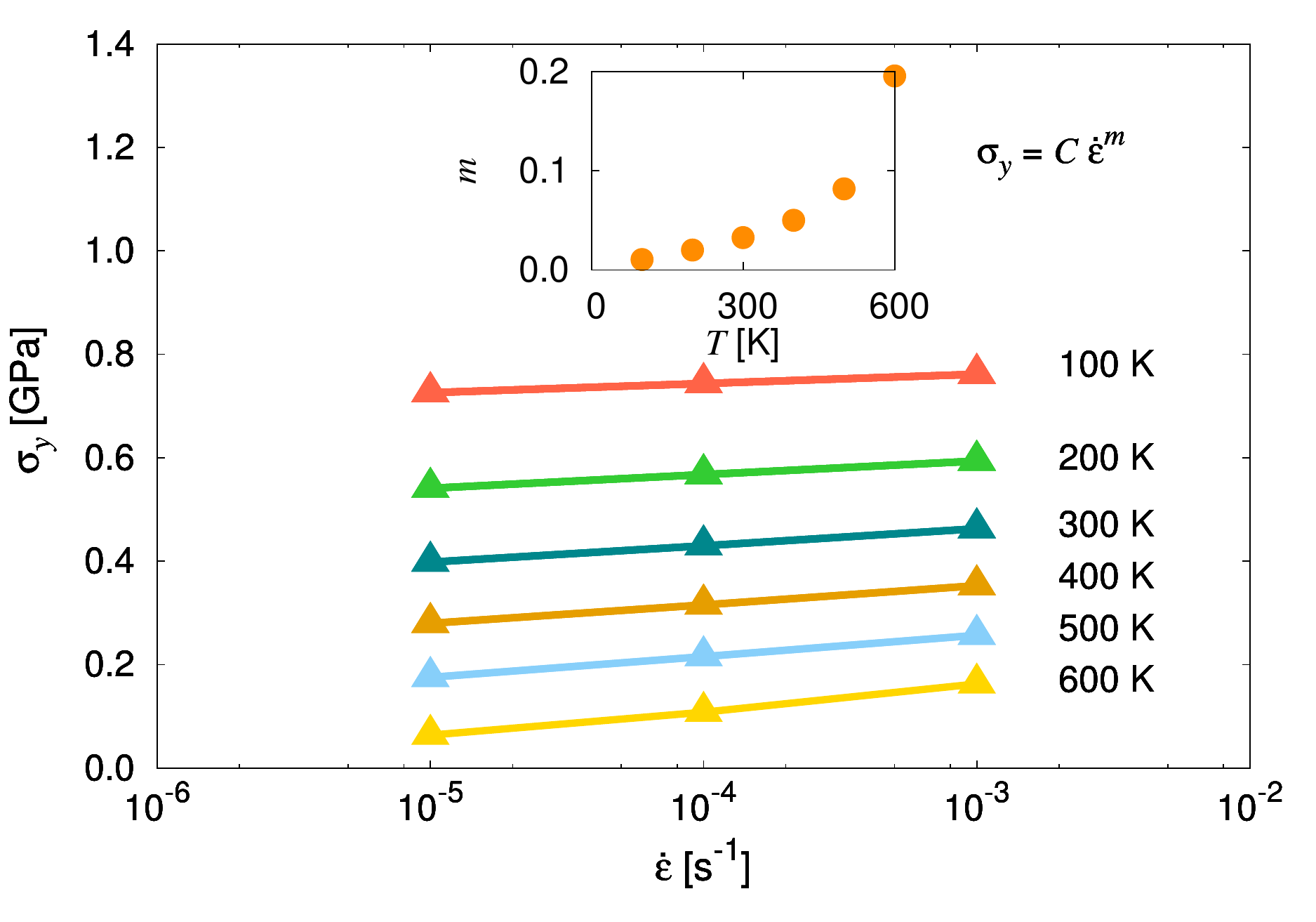}
        \caption{Dependence of yield strength with strain rate for loading along direction \hkl[101] as a function of temperature. The inset represents the dependence of the strain rate sensitivity exponent $m$ with temperature. \label{fig_yield_vs_srate}}
\end{figure}

This precisely what is done in Figure \ref{fig_yield_vs_srate} for $[101]$ loading tests. The figure shows the variation of the yield strength at the three strain rates considered here, again in the range $100<T<600$ K. The data can then be fitted to the following expression:
\begin{equation}
\sigma_y=C\dot{\varepsilon}^m
\end{equation}
where $C$ is a fitting constant. The strain rate sensitivity exponent is formally defined as:
\begin{equation}
m=\frac{\partial\log\sigma_y}{\partial\log\dot{\varepsilon}}
\end{equation}
$m$ is plotted in the inset to Fig.\ \ref{fig_yield_vs_srate}, where it can be seen that it increases monotonically with temperature from a value of $m=0.01$ at 100 K to $\approx0.2$ at 600 K. The implications of these results will be discussed in Section \ref{disc}.

\subsection{Biaxial loading tests and yield surfaces} 
\label{subsection_biaxial}

For non-associated CP formulations such as the present one, yielding is not a separate and independent criterion, but a consequence of the constitutive law of the material behavior \citep{bodner1968}. Indeed, with yielding defined on the basis of the identification criterion introduced in Section \ref{sec:yy}, yield surfaces are furnished as a product of the CP calculations. In this Section we calculate the yield curves under biaxial stress conditions for selected pairs of orthogonal loading directions $\vec{l}_{y}$ and $\vec{l}_{z}$.
As noted in Section \ref{sec:schmid}, the present implementation of the non-Schmid stress projection law is only valid for tensile conditions\footnote{Although this is not a limitation in a strict sense as it is done simply for consistency with non-Schmid treatments published in the literature.}. Thus, our yield curves are only meaningful in the positive stress quadrant (or octant, for yield surfaces). The procedure to calculate each point of the yield surface consists of deforming the system simultaneously along the prescribed orientations until the material yields on either one according to criterion \eqref{eq:crit}. The stresses $P_{zz}$ and $P_{yy}$ are then measured along both directions and the resulting duplet is added to the curve. Plane stress conditions are adopted along the remaining direction, {\it i.e.}~$P_{xx}=0$. The calculations are done at a nominal strain rate of $\dot\varepsilon=10^{-4}~{\rm s}^{-1}$, with slight variations above and below this value in one of the loading directions to accumulate different levels of stress and map the entire stress quadrant.

\begin{figure}[h]
        \centering
        \includegraphics[width = 0.7\textwidth]{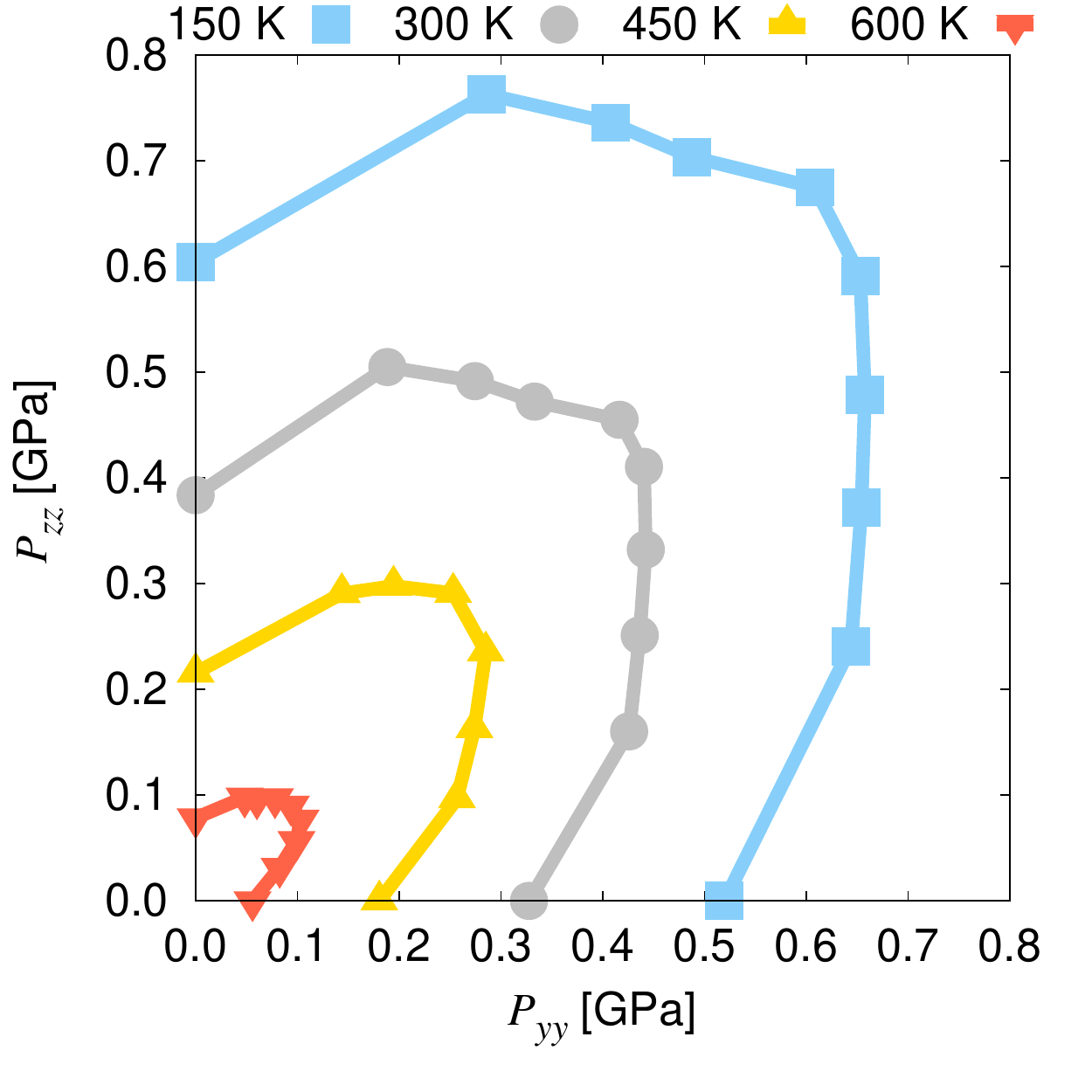}
        \caption{Yield curve for loading along directions $\vec{l}_{y}=[111]$ and $\vec{l}_{z}=$\hkl[11-2] as a function of temperature.\label{biaxial1}}
\end{figure}

First we calculate the yield curve for $\vec{l}_{y}=[111]$ and $\vec{l}_{z}=$\hkl[11-2] as a function of temperature. Results are shown in Figure \ref{biaxial1}. The curves enclose domains that are everywhere convex, thus satisfying the \emph{Drucker-Prager} criterion for stable plastic flow materials \citep{prager1952general,drucker1952extended}. The absolute values and the temperature sensitivity of the yield stresses for the end cases of $P_{zz}=0$ and $P_{yy}=0$ are consistent with the results shown in Section \ref{subsection_uniaxial} for the $\vec{l}_{y}$ and $\vec{l}_{z}$ chosen here.

\begin{figure}[h]
        \centering
        \includegraphics[trim={0cm 0cm 7cm 16cm},clip,width = 0.7\textwidth]{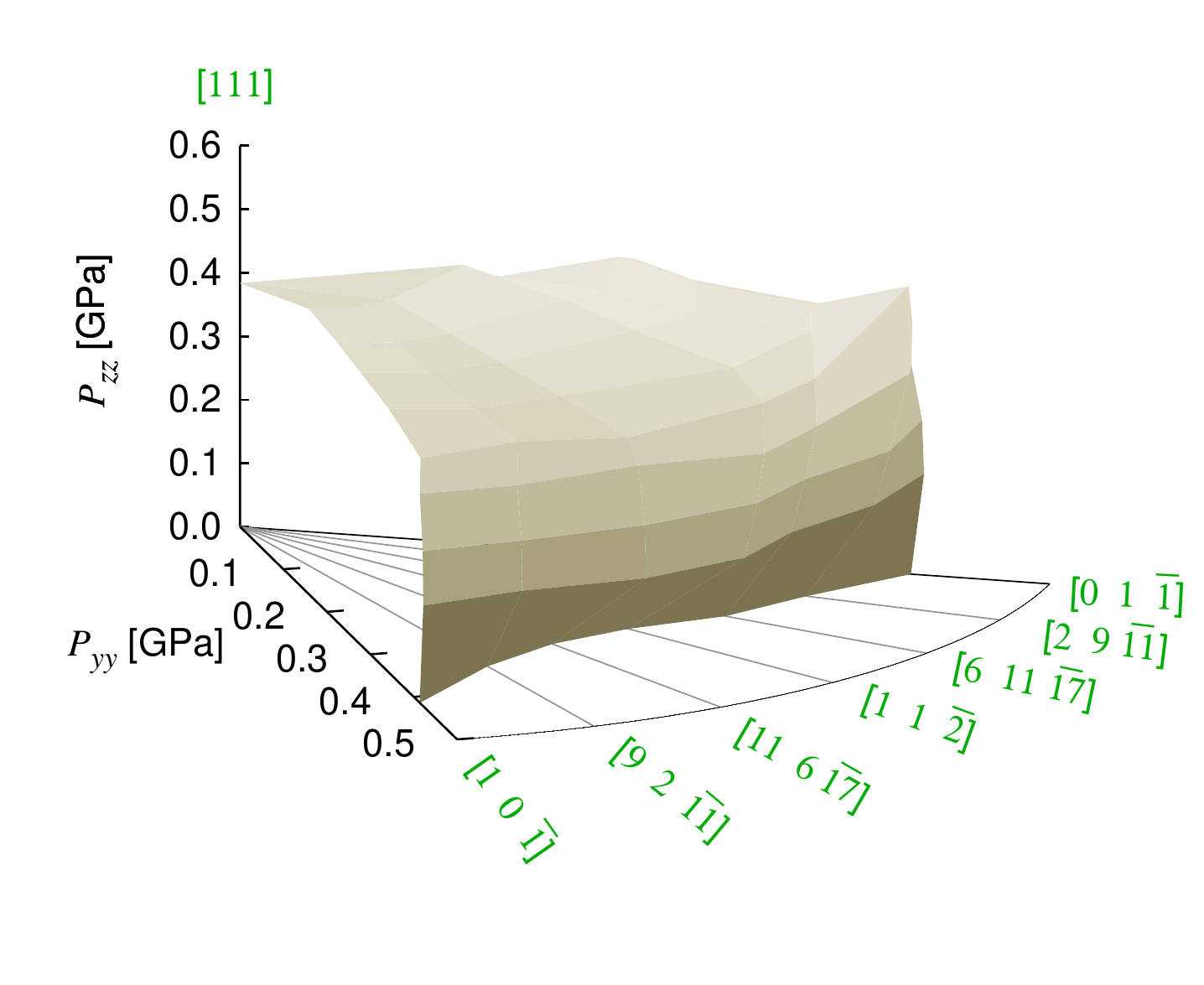}
        \caption{Yield surface at 300 K for biaxial loading along directions belonging to the $[111]$ zone. By symmetry, only the $60^{\circ}$-arc need be explored.\label{biaxial2}}
\end{figure}

The next series of calculations involves determining the entire yield surface of the $[111]$ zone, {\it i.e.}~for a set of directions orthogonal to $[111]$ in $10^{\circ}$ intervals, at a fixed temperature of 300 K. Results are shown in Figure \ref{biaxial2}. Symmetry considerations limit the angular range to be explored to a $60^{\circ}$ arc, which is shown in the figure. 
Yield surfaces such as this one are the culmination of crystal plasticity calculations, and can be used as constitutive input into continuum models to simulate effective mechanical behavior at the engineering scale, for component design and/or to simulate, {\it e.g.}, thermo-mechanical treatments \citep{sheng2004,serenelli2010}.

\section{Discussion and conclusions}
\label{disc}

In this Section we consider the most important implications of our results. First, we discuss one of the most salient characteristic of the current work. The present CP model uses a standard rate-dependent, finite-deformation, non-associated theory of crystal plasticity. However, while the underlying kinematic formulation serves as the mathematical framework upon which to build a physical methodology, it is via the connection to the material physics that the model is rendered truly predictive. Our technique does so by incorporating the following three features of bcc slip:
\begin{itemize}
\item A complete (T/AT plus nonglide) treatment of non-Schmid effects.
\item A kinematic flow rule based on a thermally-activated screw dislocation mobility.
\item Using accurate interatomic potentials for computing all the free parameters in the model.
\end{itemize}
We have shown that the full model is capable of predicting the experimentally-measured temperature dependence of yield strength in the entire temperature range for W single crystals without parameter-fitting of any kind\footnote{Of course, interatomic potentials --which form the basis of the constitutive information employed here-- are subjected to a fair amount of fitting themselves, both to experimental data and first-principles calculations. However, potential fitting is extraneous to our work, in the sense that it was neither performed by us nor done with this application in mind, while the parameter fitting that we refer to here is dedicated specifically to reproduce experimental data of interest to the application of the model.}. The sole source of material (constitutive) information is a carefully selected semi-empirical interatomic potential fitted exclusively to a DFT-generated dataset that includes the Peierls stress in its full atomistic meaning. This closes the gap seemingly separating electronic structure calculations of fundamental dislocation core properties and real measurements of the yield stress in uniaxial tensile tests of bcc materials.

Indeed, much effort has been devoted to the study of this long-standing experiment/simulation discrepancy, particularly at temperatures $<20$ K. Explanations based on collective dislocation dynamics, such as network kinetics \citep{caibulatov2002nodal} and/or mutually interacting dislocations \citep{groger2007pileup} can be more or less discounted in light of recent detailed electron microscopy experiments of isolated screw dislocation motion \citep{caillard2010_I,caillard2010_II,caillard2014}. A more  recent description, based on quantum effects at very low temperatures, has been put forward with reasonable success \citep{proville2012quantum}. On this basis, our first partial conclusion is that, while the present calculations do not provide sufficient grounds to invalidate these theories, they do clearly demonstrate that \emph{models based solely on classical mechanics --and without recourse to fitting to experimental results-- can be formulated to predict the temperature dependence of the yield strength of bcc single crystals}. Evidently, we issue this conclusion with caution, as W does not constitute by itself a representative sample to convincingly claim generality, but we believe that it constitutes a step in that direction. 

Another important physical aspect of tensile deformation in single bcc crystals is the seemingly distinct slip mechanisms operating in different temperature ranges. According to Seeger and collaborators, there are three clearly distinguishable temperature regions in the flow stress-temperature curves for bcc metals \citep{seeger1981,seeger1995flow,Brunner}, namely, the so-called upper and lower \emph{bend} temperatures, $\check{T}$ and $\hat{T}$, and the \emph{knee} temperature $T_k$\footnote{$T_k$ is understood as the temperature above which the contribution of the kink-pair formation mechanism to the flow stress becomes negligibly small, {\it i.e.}~it signals the athermal limit.}. $\check{T}$, $\hat{T}$, and $T_k$ delimit three different regimes where slip may occur on $\{110\}$, as well as $\{112\}$, glide planes, and give rise to different deformation mechanisms. Although these theories are  substantiated by ample experimental data, there are recent studies that indicate that $\{110\}$ slip may be sufficient to explain the most salient features of bcc plasticity \citep{marichal2013,ali2011}.
This is consistent with the analysis presented here, backed by atomistic input, which suggests that only $\{110\}$ slip is admissible in bcc W. Interestingly, the screw dislocation mobility law employed in this work, where $\{112\}$ slip is disallowed by construction (cf.\ Section \ref{mobb}), is sufficient to quantitatively characterize the evolution of the yield stress across the entire temperature spectrum, without any {\it ad hoc} partition of mechanisms into different temperature regimes. We emphasize once more that the screw mobility law has been fitted exclusively to first-principles data.

In Section \ref{mmmm} we have provided calculations of the strain rate sensitivity defined as $m=\partial\log\sigma_y/\partial\log\dot{\varepsilon}$. It must be noted that our value of $m=0.023$ at 300 K obtained in the $10^{-3}>\dot{\varepsilon}>10^{-5}$ s$^{-1}$ range is consistent with measurements performed by \citet{zurek1991} in W compressed uniaxially at strain rates from $10^{-3}$ to $10^{3}$ s$^{-1}$. Notwithstanding the differences in experimental methodology and strain rate regime, this is also encouraging agreement for a result other than yield. $m$ is an important parameter for calculating the kink-pair activation enthalpy and activation volume from stress-relaxation tests. Note that some authors use an alternative definition for the strain rate sensitivity \citep{raffo1969,Brunner}, namely, $\lambda=\partial\sigma/\partial\log\dot{\varepsilon}$, which is related to $m$ via $\lambda=m\sigma$. We can then conclude that \emph{the agreement achieved for a derivative quantity of the yield stress such as $m$ is symptomatic of the quality of the method outside the primary validation space}.

The advantages of this and other CP methodologies w.r.t.~more accurate techniques such as molecular dynamics, dislocation dynamics, or phase field methods is of course their computational expediency. Backed by the encouraging outcome of the validation exercise, this has enabled us to map the entire loading orientation space in the standard triangle (231 directions) as a function of temperature in a experimentally-meaningful strain rate range. These results can then be used to extract useful information, such as the strongest and softest orientations as a function of temperature and strain rate, or the strain rate sensitivity of our W model system. This information can ultimately be used to define yield criteria under a variety of conditions for more homogenized methods, with the aim put on component design.

In this sense, the culmination of the CP simulations is the calculation of yield curves and yield surfaces in stress space. The stress space that we have chosen for our yield surface calculations is a purely biaxial one (in plane stress) with one fixed direction, chosen arbitrarily to be $[111]$, and the family of orthogonal directions taken in $10^{\circ}$ intervals. This biaxial loading configuration is the elementary basis for pressurized cylinders, {\it e.g.}~pipes, and is thus useful to design components based on this geometry. As well, it can serve as the design premise for loaded plates under plane stress conditions. 
It is of interest to note that yield surfaces can also serve as the plastic potential in the fundamental theory of plasticity \citep{lubliner_book}. This equivalence is valid when the critical resolved shear stress is not dependent on the current stress state\footnote{Particularly on non-deviatoric components.} \citep{lubliner_book,starovoitov2012}. However, this may not be applicable in the present model, where the CRSS is seen to display a strong dependence on hydrostatic (nonglide) stress components as discussed in Section \ref{sec:schmid}. This is also the case in rock and soil plasticity ({\it e.g.}~\citet{pariseau1968}). In such cases, the normality rule is referred to the pressure-dependent yield surface instead.

A standing limitation of our model is that we have only made use of the tensile region of the dependence of the critical stress $\tau_c^{\chi}$ with the nonglide stress $\sigma$ (cf.\ eq.\ \eqref{tauc}). Of course, this dependence is essential to characterize the tension/compression asymmetry customarily observed in bcc crystals, cf.\ Section \ref{intro}. However, this is only a weak limitation, as the present CP formulation is sufficiently flexible to admit a full (nonlinear) fit to the data shown in Fig.\ \ref{non}.
Finally, we emphasize that the present study focuses on plastic yielding, and consequently, we have not explored the evolution of the flow stress much beyond the extent needed to define a robust yield criterion (cf.\ Section \ref{sec:yy}). However, this does not detract from the validity of the dislocation density evolution model presented in Section \ref{evol}, which has been used prolifically in many CP studies (cf.\ Section \ref{intro}), and which is being investigated in ongoing studies.

\section*{Acknowledgments}
This work was performed under the auspices of the U.S. Department of Energy by Lawrence Livermore National Laboratory under Contract No.\ DE-AC52-07NA27344. J. M. acknowledges support from DOE's Early Career Research Program. 
D. C. acknowledges support from the Consejo Social and the PhD program of the Universidad Polit\'ecnica de Madrid. The authors are indebted to Dr. M. Victoria for inspiration, encouragement, and guidance throughout this work.

\clearpage
\appendix

\section{$\{110\}\langle111\rangle$ slip systems and latent hardening matrix considered for bcc W.}
\label{app:slip}

\begin{table}[h]
\caption{Slip systems considered in our calculations, listing the non-normalized crystallographic vectors $\vec{m}^\alpha$,$\vec{n}^\alpha$ and $\vec{n}_1^\alpha$. Note that in DAMASK each slip system is taken both in its positive and negative sense, which is equivalent to formulations where 24 positive slip systems are employed \citep{groger2008_II}.}
\begin{center}
\begin{tabular}{ccccc}
\hline
$\alpha$ & Reference system &  $\vec{m}^\alpha$ & $\vec{n}^\alpha$ & $\vec{n}_{1}^\alpha$ \\[0.5ex]
\hline
1  &	 \hkl[1-11]  \hkl(011)	& \hkl[1-11]  &	\hkl[011]  & \hkl[-101] 	\\[0.25ex]
2  &	 \hkl[-1-11] \hkl(011)  	& \hkl[-1-11] & \hkl[011]  & \hkl[-1-10] \\[0.25ex]
3  &	 \hkl[111] 	 \hkl(0-11)	& \hkl[111] 	  & \hkl[0-11] & \hkl[1-10] 	\\[0.25ex] 
4  &	 \hkl[-111]  \hkl(0-11)	& \hkl[-111]  &	\hkl[0-11] & \hkl[101] 	\\[0.25ex]
5  & \hkl[-111]  \hkl(101)	& \hkl[-111]  &	\hkl[101]  & \hkl[110] 	\\[0.25ex]
6  &	 \hkl[-1-11] \hkl(101)	& \hkl[-1-11] & \hkl[101]  & \hkl[011] 	\\[0.25ex]
7  &	 \hkl[111]   \hkl(-101)	& \hkl[111]   & \hkl[-101] & \hkl[0-11]  \\[0.25ex]
8  &	 \hkl[1-11]  \hkl(-101)	& \hkl[1-11]  &	\hkl[-101] & \hkl[-1-10] \\[0.25ex]
9  &	 \hkl[-111]  \hkl(110) 	& \hkl[-111]  &	\hkl[110]  & \hkl[01-1]	\\[0.25ex] 
10 & \hkl[-11-1] \hkl(110) 	& \hkl[-11-1] & \hkl[110]  & \hkl[10-1]  \\[0.25ex]
11 & \hkl[111]   \hkl(-110)	& \hkl[111]   & \hkl[-110] & \hkl[-101]  \\[0.25ex]
12 &	 \hkl[11-1]  \hkl(-110)  & \hkl[11-1]  & \hkl[-110] & \hkl[011]  \\[0.25ex]
\hline
\end{tabular}
\end{center}
\label{tabslip}
\end{table}%


\begin{table}[h]
\caption{Interaction coefficients $\xi_{\alpha\alpha'}$ for the 12 slip systems defined in Table \ref{tabslip}. The letter coding employed is `A': self; `CP': coplanar; `CL': collinear; `O': orthogonal; `G': glissile; `S': sessile. The reader is referred to Table \ref{table2} for the numerical value of each coefficient.}
\begin{center}
\begin{tabular}{|c|cccccccccccc|}
\hline
$\alpha$ & 1 & 2 & 3 & 4 & 5 & 6 & 7 & 8 & 9 & 10 & 11 & 12 \\
\hline
1  &  A  &&&&&&&&&&&\\ 
2  &  CP & A  &&&&&&&&&&\\
3  &  S  & S  & A  &&&&&&&&&\\
4  &  S  & S  & CP & A  &&&&&&&&\\
5  &  G  & O  & O  & CL & A &&&&&&&\\
6  &  O  & CL & G  & O  & CP & A  &&&&&&\\
7  &  O  & G  & CL & O  & S  & S  & A  &&&&&\\
8  &  CL & O  & O  & G  & S  & S  & CP & A  &&&&\\
9  &  O  & G  & O  & CL & CL & O  & G  & O  & A  &&&\\
10 &  CL & O  & G  & O  & O  & G  & O  & CL & CP & A &&\\
11 &  G  & O  & CL & O  & G  & O  & CL & O  & S  & S & S  &\\
12 &  O  & CL & O  & G  & O  & CL & O  & G  & S  & S & CP & A \\
\hline
\end{tabular}
\end{center}
\label{tabxi}
\end{table}%

\clearpage
\section{Details on the atomistic calculations of non-Schmid parameters}
\label{app_nonschmid}

Critical stresses are computed by applying shear stresses incrementally to a simulation box containing a screw dislocation lying on a glide plane forming an angle $\chi$ with the MRSS plane. The system is schematically shown in Fig.\ \ref{davinci}. The box dimensions vary slightly with orientation, such that, for $\chi=0$, the box contains 3024 atoms and the dimensions are $21a\times24b\times1c$, where $a$, $b$, and $c$ are the moduli of the bcc lattice vectors $x\equiv[\bar{1}21]$, $y\equiv[\bar{1}01]$, and $z\equiv[111]$, respectively. The calculations are performed using the \emph{nudged elastic band} (NEB) method \citep{NEB_Henkelman2000} implemented in the parallel molecular dynamics code LAMMPS \citep{lammps}. Periodic boundary conditions are applied along the dislocation line direction $z$ while non-periodic and shrink-wrapped boundary conditions are applied along the $y$ and $x$ directions. The transition path selected for the NEB calculations is a linear trajectory along the reaction coordinate joining two consecutive Peierls valleys, where the dislocation is relaxed to equilibrium.


Three different forces are applied to different groups of atoms in the simulation box in order to calculate $\sigma^c_{\chi}$. These forces recreate the stress tensor \eqref{tensor} in the simulation box: 
\begin{enumerate} 
\item First, an external force $f_{z}$ is added to the atoms on the boundary surfaces of the simulation box perpendicular to the $y$-axis to study the T-AT asymmetry. The external force per atom is $f_{z}= \frac{\tau L_{x} L_{z}}{N_{z}}$, where $\tau$ is the desired shear stress, $N_{z}$ is the number of atoms in each nonperiodic surface along $z$ and $L_{x} L_{z}$ is the cross-sectional area of the each of the bounding surfaces along to $y$. 
\item To study the contribution from nonglide stresses, an external force $f_{x}$ is added to the atoms on the boundaries of the simulation box perpendicular to the $x$-axis. The external force per atom is obtained as $f_{x}= \frac{\sigma L_{y} L_{z}}{N_{x}}$, where $\sigma$ is the applied nonglide stress, $N_{x}$ is the number of atoms in each surface and $L_{y}L_{z}$ is the cross-sectional area of the each of the  surfaces along $x$.
\item Further, an external force $f_{y}$ is added to the atoms on the surfaces along the $y$ direction, additionally to the shear stress $\tau$. $f_{y}$  is defined as $f_{y} = \frac{\sigma L_{x} L_{z}}{N_{y}}$, with $N_{y}=N_z$ and $L_{x}L_{z}$ is the area of the each of the surfaces perpendicular to \textit{z}.
\end{enumerate}
31 intermediate replicas are used in the NEB calculations to capture the trajectory and measure the critical stress.

\clearpage
\section{Crystal plasticity calculations of flow stress dependence with orientation and temperature}
\label{app_flow}

\begin{figure}[h]
\centering
\includegraphics[width=0.85\linewidth]{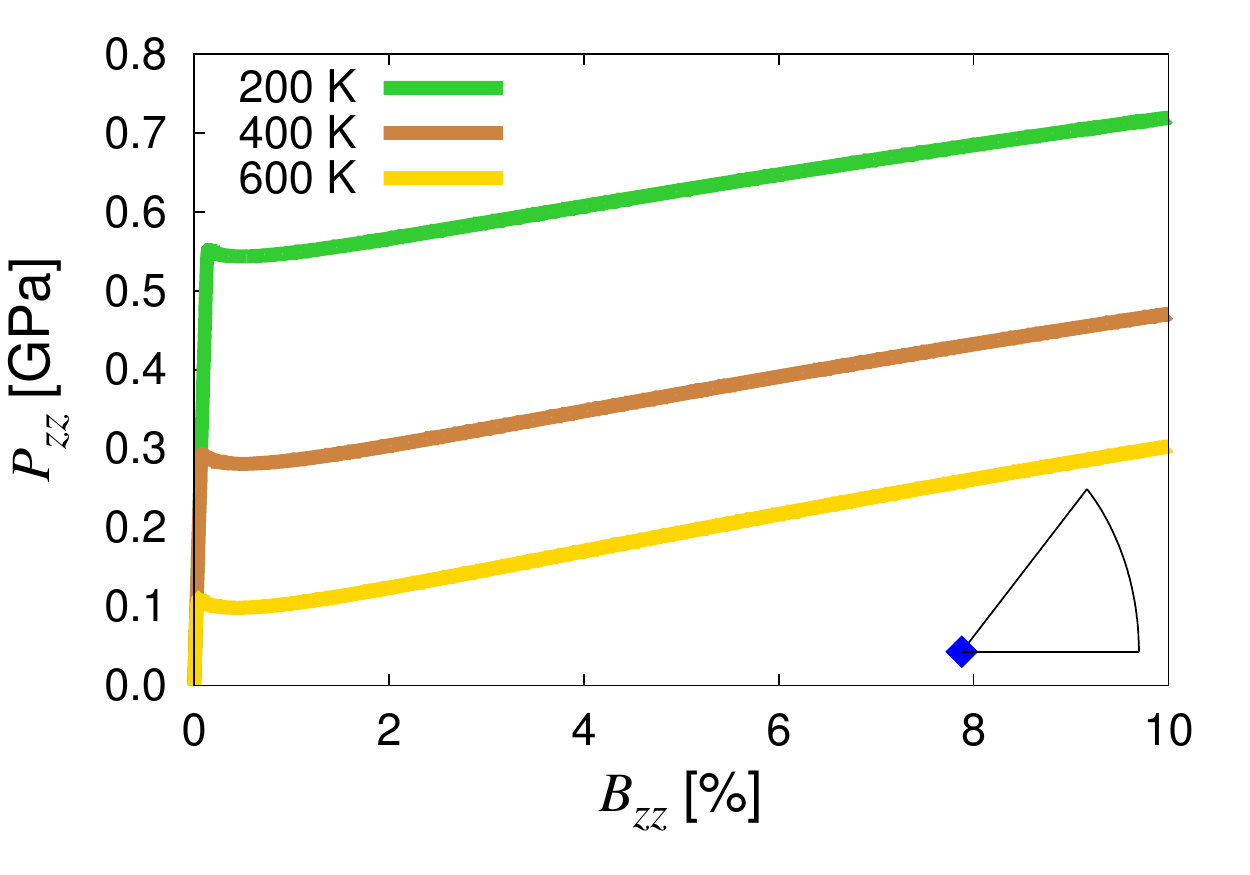}
\caption{Stress-strain curves for uniaxial loading along the [001] orientation at a strain rate of $10^{-3}$ s$^{-1}$ for three different temperatures. These curves are representative of multi-slip conditions where Taylor-type hardening is enabled\label{flowst1}.} 
\end{figure}

To demonstrate the performance of the model in the flow stress regime, we carry out calculations for a few selected orientations and temperatures up to $10\%$ strain. Figure \ref{flowst1} shows the stress-strain response at a strain rate of $10^{-3}$ s$^{-1}$ as a function of temperature for the $[001]$ loading orientation. This is an orientation conducive to multi-slip and thus the system is expected to harden in accordance with eqs.\ \eqref{fftt}, \eqref{lpp}, and \eqref{prime} as the deformation progresses. The figure shows results for the full non-Schmid model.
For general viscoplastic materials it is common to represent the $\sigma$-$\varepsilon$ relation as a power law of the type:
\begin{equation}
\sigma=K\varepsilon^n
\label{hard1}
\end{equation}
where $K$ is a constant and $n$ is the so-called \emph{hardening exponent}. Accordingly, the hardening rate can be expressed as:
\begin{equation}
\frac{d\sigma}{d\varepsilon}=Kn\varepsilon^{n-1}
\label{hard2}
\end{equation}
Fits of eq.\ \eqref{hard1} to the data in Fig.\ \ref{flowst1} yield values of $n=0.82$, 0.86, and 0.87 for $T=200$, 400, and 600 K, respectively. From eq.\ \eqref{hard2}, these numbers result in hardening rates of $\frac{d\sigma}{d\varepsilon}\approx20$ MPa/\% in all cases.

\begin{figure}[h]
\centering
	\subfigure[\text{[111] loading}]{
			\includegraphics[width=0.7\linewidth]{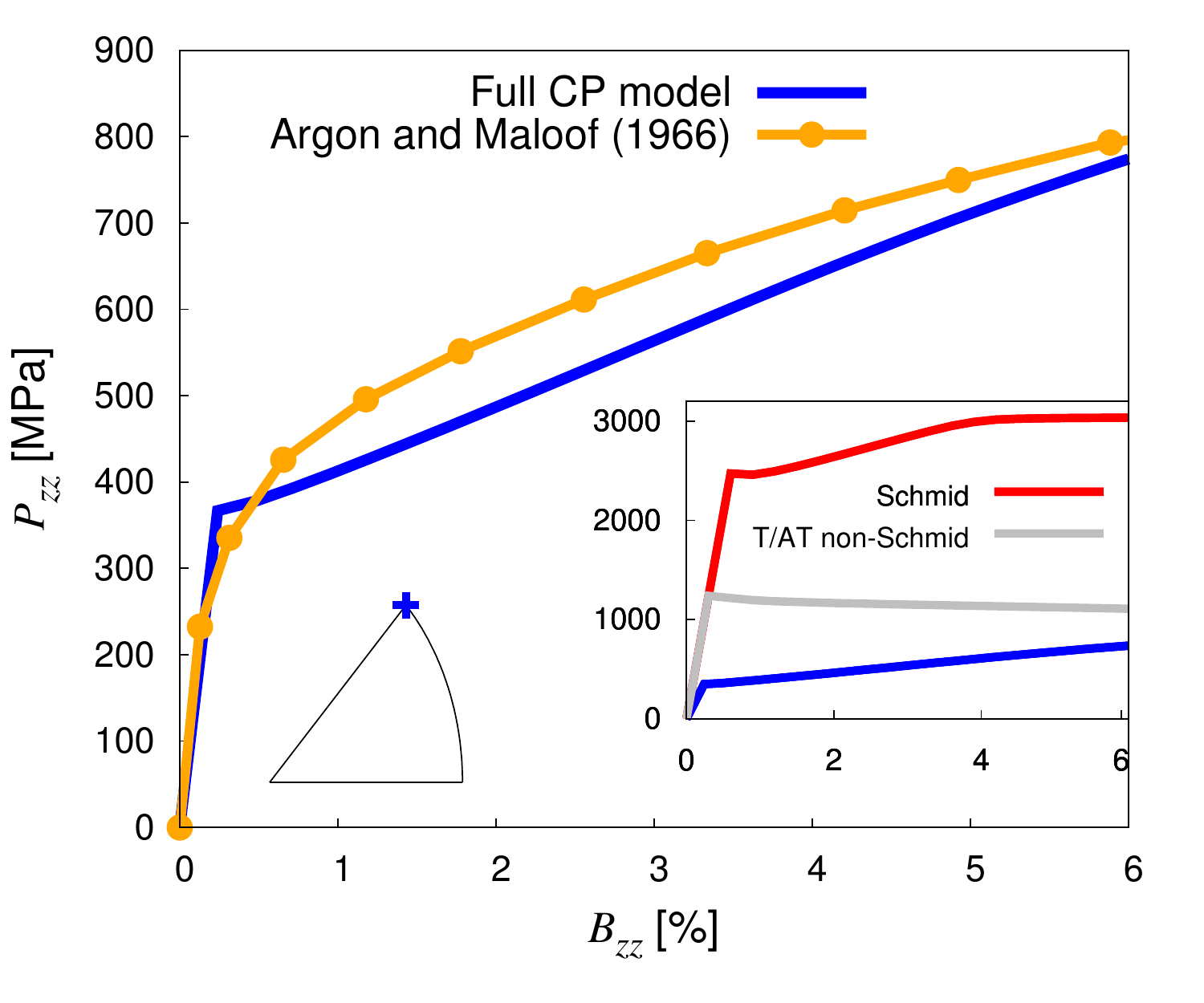}
 \label{a111}}
    	\subfigure[\text{[110] loading}]{
        \includegraphics[width=0.7\linewidth]{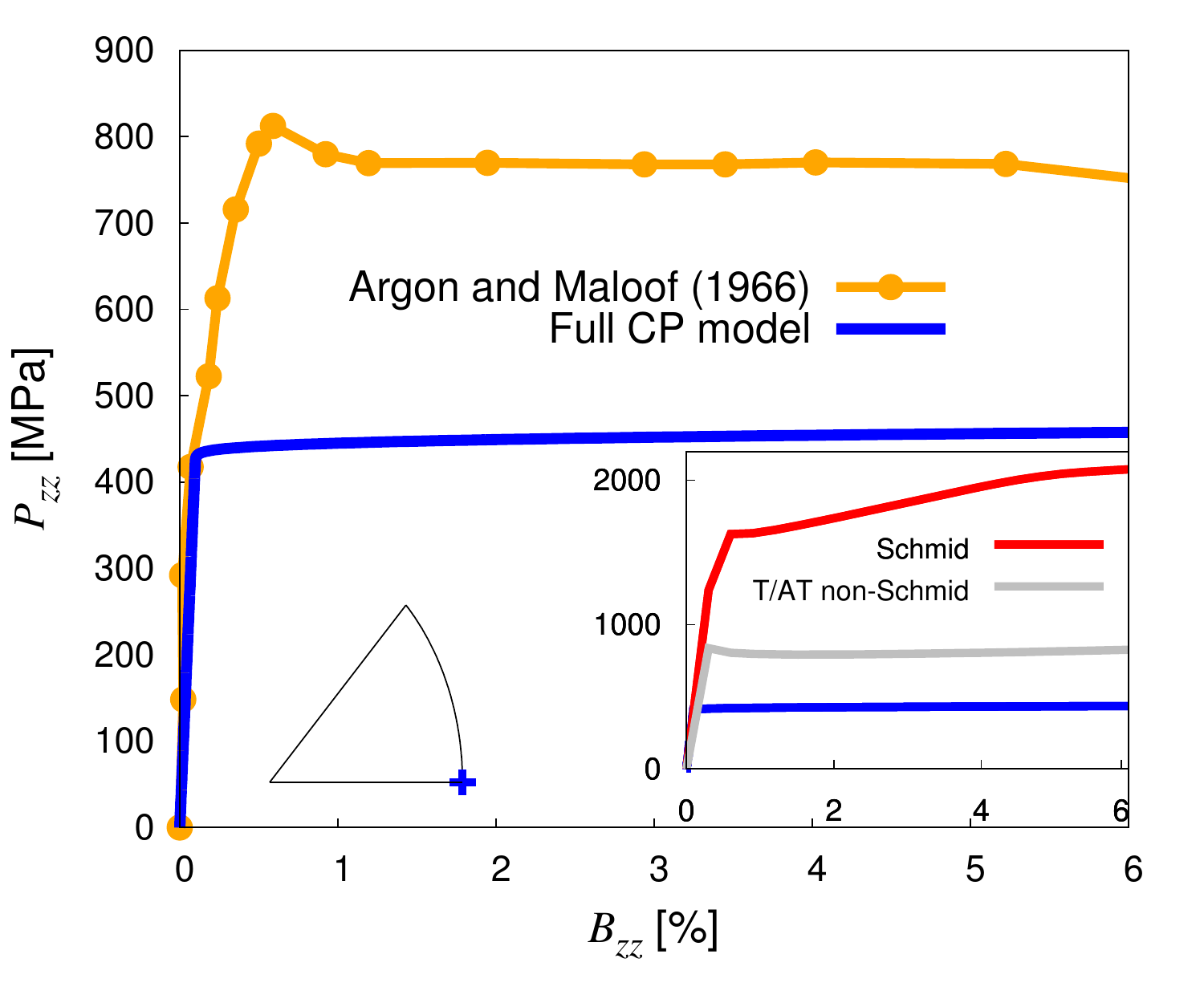}
    	\label{a101}}
	   \caption{\label{flowstr2}Flow stress of W single crystals at the conditions used by \citet{argon_maloof_1966} (cf.\ Section \ref{validation}) in tensile deformation tests under two different loading orientations. The experimental data is shown for comparison. The inset shows the results of CP calculations with different contributions of the projection tensor activated.}
\end{figure}

Next, we compare the model against the experimental results of \citet{argon_maloof_1966} for [111] and [110] loading at $\dot{\varepsilon}=10^{-4}$ s$^{-1}$. To avoid comparing in conditions where twinning may be operative ($<200$ K), which is not captured by our model, we carry out simulations at 293 K. The results are shown in Figure \ref{flowstr2}, which reveals a good agreement between the full non-Schmid model and the experimental data in the [111] loading case. According to \citet{argon_maloof_1966}, yielding under [110] loading occurs at approximately 460 MPa, which is immediately followed by an abrupt hardening stage that plateaus at $\varepsilon\approx0.8~\%$  to a value of $\approx760$ MPa. Whether or not this is the case, this initial hardening period is not captured by our model.
Under both loading conditions, however, the model is seen to reproduce the hardening rates in close agreement with the experimental data.

We emphasize that the results shown in Fig.\ \ref{flowstr2} have been obtained without fitting to experimental (or otherwise) stress-strain curves of any kind, and so the model appears to capture the essential features of plastic flow for the selected conditions showcased here. As noted earlier, these preliminary results do not imply that the model is suitable for calculating the flow stress under general loading conditions.



\clearpage
\bibliographystyle{model5-names}
\bibliography{biblio_cp_disloucla_vs5.bib}





\end{document}